\documentclass{article}

\usepackage[letterpaper,top=2cm,bottom=2cm,left=3cm,right=3cm,marginparwidth=1.75cm]{geometry}

\usepackage{amsmath}
\usepackage{graphicx}
\usepackage{graphicx}
\usepackage{dcolumn}
\usepackage{bm}

\usepackage{graphicx} 
\usepackage{array}
\usepackage{multirow}
\usepackage{booktabs}
\usepackage{rotating}
\usepackage{tabu}
\usepackage{relsize}
\usepackage{amssymb}
\usepackage{pifont}
\usepackage{bbding}
\usepackage{starfont}
\usepackage{wasysym}
\usepackage{eqnarray}
\usepackage{tablefootnote}
\usepackage{ragged2e}
\usepackage[numbers]{natbib}
\usepackage{xcolor}
\usepackage{amsmath}
\usepackage{rotating}

\definecolor{indigo1}{rgb}{0 , 0.4470,  0.7410}
\definecolor{orange1}{rgb}{0.8500 0.3250 0.0980}
\definecolor{yellow1}{rgb}{0.9290, 0.6940, 0.125}
\definecolor{purple1}{rgb}{0.4940, 0.1840, 0.5560}
\definecolor{green1}{rgb}{0.4660, 0.6740, 0.1880}
\definecolor{blue1}{rgb}{0.3010, 0.7450, 0.9330}
\definecolor{crimson1}{rgb}{0.6350, 0.0780, 0.1840}
\definecolor{lavender1}{rgb}{0.996, 0.812, 0.996}
\definecolor{brown1}{rgb}{0.6050, 0.4023, 0.2344}
\definecolor{teal1}{rgb}{0, 0.847, 0.847}
\definecolor{darkblue1}{rgb}{0, 0.0664, 0.6055}
\definecolor{peach1}{rgb}{1, 0.69, 0.486}
\definecolor{gray1}{rgb}{0.8, 0.804, 0.776}
\definecolor{darkgreen1}{rgb}{0.349, 0.506, 0.322}
\definecolor{red1}{rgb}{0.89, 0.141, 0.169}

\title{Axisymmetric displacement flows in fluid-driven fractures}
\author{Sri Savya Tanikella$^1$, Emilie Dressaire$^{1}$\\
1. University of California Santa Barbara}

\begin{document}
\maketitle

\begin{abstract}
Displacement flows are common in hydraulic fracturing, as fracking fluids of different composition are injected sequentially in the fracture. The injection of an immiscible fluid at the center of a liquid-filled fracture results in the growth of the fracture and the outward displacement of the interface between the two liquids. We study the dynamics of the fluid-driven fracture which is controlled by the competition between viscous, elastic, and toughness-related stresses. We use a model experiment to characterize the dynamics of the fracture for a range of mechanical properties of the fractured material and fracturing fluids. We form the liquid-filled pre-fracture in an elastic brittle matrix of gelatin. The displacing liquid is then injected. We record the radius and aperture of the fracture, and the position of the interface between the two liquids. In a typical experiment, the axisymmetric radial viscous flow is accommodated by the elastic deformation and fracturing of the matrix. We model the coupling between elastic deformation, viscous dissipation, and fracture propagation and recover the two fracturing regimes identified for single fluid injection. For the viscous-dominated and toughness-dominated regimes, we derive scaling equations that describe the crack growth due to a  displacement flow and show the influence of the pre-existing fracture on the crack dynamics through a finite initial volume and an average viscosity of the fluid in the fracture.
\end{abstract}

\section{Introduction}
Fluid-driven or hydraulic fracturing results from the injection of a pressurized fluid in low permeability solid media. The formation and propagation of the fluid-filled tensile fracture is commonly observed in engineering and natural geophysical processes. For example, the formation of magma-driven dykes is due to density differences that generate pressure large enough to propagate a vertical fracture in the surrounding rock \citep{Lister1991, Rubin1995}. The most common industrial application of hydraulic fracturing is well stimulation to facilitate the extraction of oil and gas from shale formations \citep{Cueto2013}. A fluid is injected at high pressure to expand fractures initiated with small-scale explosions in unconventional reservoirs. The fractures constitute new flow pathways, facilitating fluid transport and storage in low permeability and low porosity rock formations. Other applications leverage the enhanced transport. For example, fractures connecting wells can be used to extract geothermal energy as the fluid pumped through the fracture heats up as it travels underground \citep{Murphy1981, Caulk2016, Luo2017}. Fractures are also used for storage, including carbon sequestration \citep{Huppert2013, Jia2019} and disposal of liquid waste \citep{Bao2016, Alessi2017}.

 The complex mechanics of fluid-driven fractures are controlled by the deforming boundary, fluid flow and stress singularity  at the tip of the tensile fracture \citep{Detournay2016}. Field testing, laboratory-scale experiments, and predictive modelling evidence that the fracture propagation is characterized by multiple length and time scales. When a Newtonian fluid is injected from a point source, in an infinite, homogeneous, and impermeable medium, a single fracture propagates radially. The elastic stress in the medium leads to the growth of a penny-shaped fluid-filled fracture in the direction of minimum confining stress. The propagation of such fractures has been studied extensively as it is essential to the modelling of more complex geological situations, including those involving a finite medium \citep{Bunger2005}, complex fluids \citep{Barbati2016, Hormozi2017, Lai2018}, and interacting fractures \citep{OKeeffe2018b}. Seminal work on the stress distribution in a penny-shaped fracture \citep{sneddon1946} and the injection of viscous fluids to form fractures \citep{Khris1955, Barenblatt1956} led to the development of self-similar solutions for fractures whose propagation is limited by the viscous dissipation in the fluid \citep{Spence1985}. Further work on the vicinity of the crack front or tip region identified two asymptotic regimes for the tip geometry and fracture propagation \citep{Desroches1994, Garagash2000, Detournay2003, Savitski2002, Garagash2005}. In the viscous-dominated scaling, the viscous dissipation in the flow opposes the elastic stress of the deformable boundary to control the fracture evolution. In the toughness-dominated regime, the material toughness opposes the elasticity-driven propagation of the fracture and determines the system's behaviour. Laboratory-scale experiments commonly rely on clear brittle elastic gels to study the crack tip region  and the penny-shaped fracture \citep{Takada1990, Kavanagh2013, Giuseppe2009, Baumberger2020}. Injections of water, glycerol, and oil in gelatin and polyacrylamide have validated the existence of two propagation regimes and the corresponding scaling laws  \citep{Lai2015, Lai2016a, OKeeffe2018a}. The behaviour of the crack tip region was studied by injecting liquid between two plates of  polymethylmethacrylate (PMMA) glued by an adhesive \citep{Bunger2008}.

 Most hydraulic fracturing processes involve multiphase flows and in particular, displacement flows \citep{Hormozi2017, Osiptsov2017, Lai2018, Wang2018, Bess2021}. During hydraulic fracturing operations several fluids are injected ranging from low-viscosity fluids to high-viscosity polymer solutions \citep{Moukhtari2018, Bess2021} and proppant slurries \citep{Hormozi2017, Wang2018, Barboza2021}. This sequence of injections aims at increasing the surface area of the fracture and at keeping the fracture open during the hydrocarbon extraction. For example, carbon dioxide injection is a promising strategy to enhance oil recovery after primary production of shale oil reservoirs \citep{Huppert2013, Jia2019}. The rapid injection of supercritical CO$_2$ in water-filled fractures is followed by the slower permeation of CO$_2$ into the rock and the migration of the oil into the fracture. This strategy increases the amount of oil recovered while storing carbon in the rock. Finally enhanced geothermal systems rely on fractures in hot rocks to connect the injection and extraction wells \citep{Murphy1981, Caulk2016, Parisio2020}. Fracturing fluids are injected first, and then the working fluid, commonly water or CO$_2$ is pumped into the fracture to extract heat.  

 \begin{figure}
  \centerline{\includegraphics[scale=0.23]{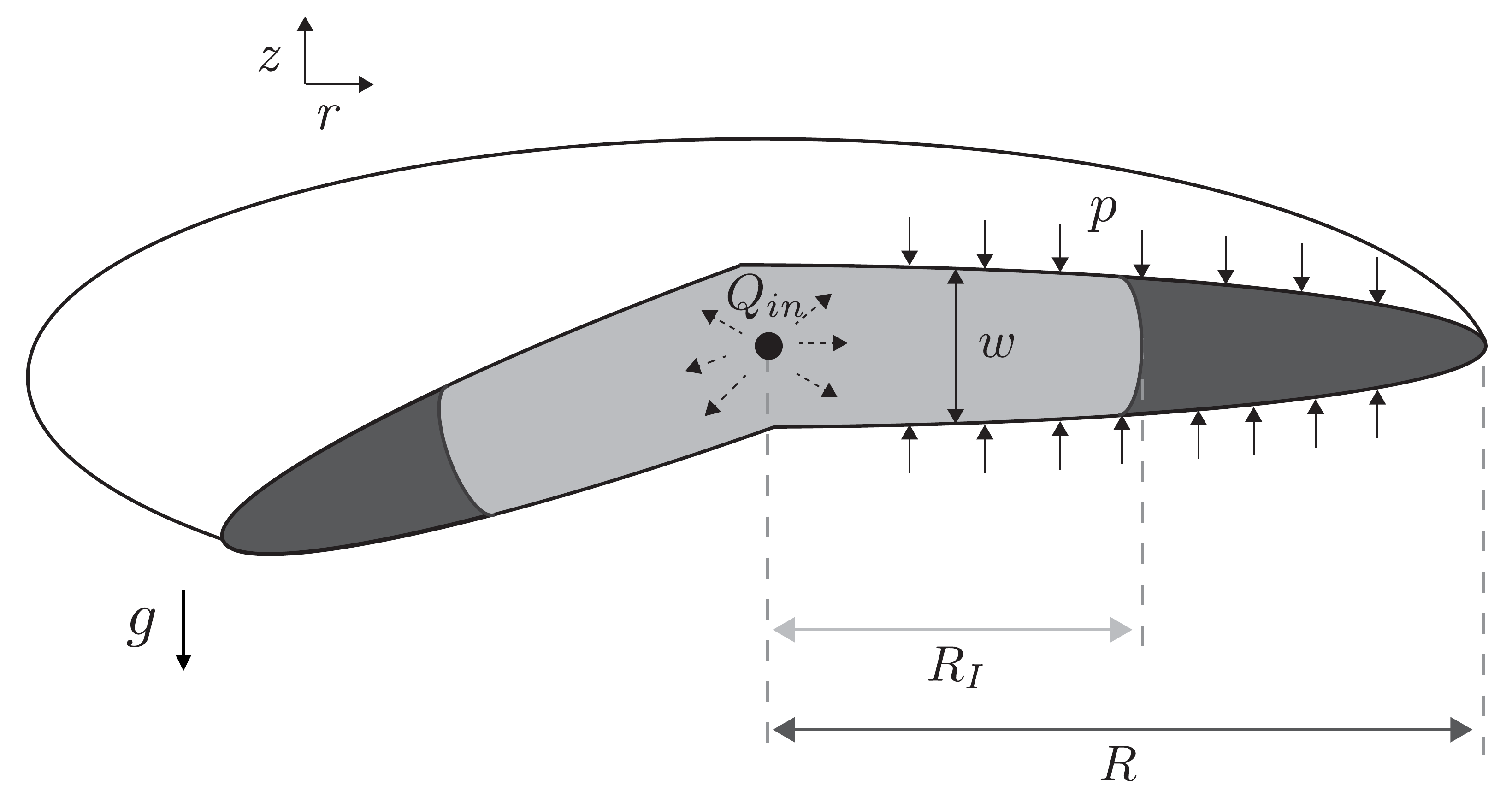}}
  \caption{Schematic of a penny-shaped fracture formed by successively injecting two liquids, first the outer fluid (dark gray) and then the inner fluid (light gray). The fracture is axisymmetric.}
\label{fig:ka}
\end{figure}

Displacement flows in porous media can give rise to complex out-of-equilibrium flow patterns when the invading fluid has a lower viscosity than the fluid which occupies the porous medium, and is referred to as the displaced or defending fluid. Practically, the patterns generated by liquid-liquid or gas-liquid displacement flows lead to preferential flow pathways in the porous medium. Extensive work has therefore been dedicated to the formation and geometry of the patterns, ranging from experimental to numerical and theoretical \citep{Saffman1958, Paterson1981, Park1984, Chen1989, Homsy1987,Lenormand1988, Tanveer1993, Primkulov2019}.  Viscous and capillary forces can contribute to fluid/fluid displacement in a porous medium. A displacement flow is characterized by two dimensionless parameters: the viscosity or mobility ratio $M=\mu_{inv}/\mu_{def}$ and the capillary number of the invading fluid $Ca = {\mu_{inv} u}/{\gamma}$ where $\mu_{inv}$ and $\mu_{def}$ are the viscosities of the invading and defending fluids, respectively, $u$ is the characteristic velocity and $\gamma$ is the surface tension of the interface between the two fluids. The influence of the two dimensionless parameters on the geometry of the invading front is summarized in Lenormand's phase diagram \citep{Lenormand1988}, which was recently revisited by \citet{Primkulov2021} to include wettability. In summary, for large Ca, the viscous forces control the system dynamics and the interfacial forces are negligible. If the invading fluid is more viscous than the defending fluid ($M\geq 1$), a compact front or interface moves through the porous medium. If the invading fluid is less viscous than the defending one ($M<<1$), the Saffman-Taylor instability leads to an unstable front with the formation of a viscous fingering pattern, observed in porous media of different complexity, from Hele shaw cells \citep{Saffman1958, Tabeling1987} to intricate networks of pores and throats \citep{Lenormand1983, Zhao2016}. For low Ca, the interfacial forces contribute to the system dynamics and result in more complex patterns at the interface between the two fluids, depending on the local pore geometry and wettability \citep{Stokes1986, Cottin2010}. The rich dynamics of displacement flows is reported in various model porous media, including networks of microchannels and rough fractures \citep{Glass2003, Chen2017, Yang2019}. Yet, the system geometry can delay the onset of viscous fingers and even suppress the Saffman-Taylor instability. In a converging Hele-Shaw cell, the stability of the interface depends on the mobility ratio, but also the characteristic velocity, the gradient of cell depth, and the contact angle at the interface \citep{Al2012, Al2013, Lu2019}. Below a critical capillary number, a compact front is observed in a converging Hele-Shaw cell despite the unfavourable nature of the displacement. Similar results are reported in flexible cells, whose geometry depends on elastohydrodynamic interactions, such as displacement flows under elastic membranes \citep{Pihler2012,Pihler2013,Pihler2015,Peng2015}. Two physical mechanisms contribute to the stabilization of the interface under an elastic membrane that deforms as fluid is injected. Firstly, the local increase in cell depth leads to a depth gradient which has been shown to delay viscous fingering for rigid converging cells. Secondly, the increase in depth reduces the characteristic velocity or capillary number corresponding to a given injection flow rate. 

 The purpose of the present paper is to model axisymmetric two-phase flows in fluid-filled growing fractures. Experiments and theoretical modelling focus on immiscible two-phase flows with a mobility ratio smaller than or of order 1 and a low capillary number, ensuring the propagation of a compact front. We build on the approach of \citet{Savitski2002} to study the coupling between the two-phase flow and the fracture growth, in the viscous and toughness regimes ($\S\,2$). We derive new scalings for the radius and aperture of the fracture and the position of the interface for immiscible displacement flows in elastic, brittle, and impermeable media ($\S\,3$). To test the scalings, we conduct injection experiments in gelatin, which is a common model medium ($\S\,4$). During the two consecutive injections, we record the geometrical parameters of the fracture and compare their time dependence with scalings ($\S\,5$). Finally, we discuss the timescales of the fracturing displacement flow ($\S\,6$).

 \section{Theoretical models}
 In this section, we model the displacement flow that is responsible for the propagation of a crack during the successive injections of two immiscible fluids. The mathematical models presented build on the framework originally introduced by \citet{Spence1985} and further developed in recent studies of single fluid injection \citep{Savitski2002, Lai2015, OKeeffe2018a}. Past work has focused on the injection of a single incompressible fluid in an elastic brittle solid through a point source (see figure \ref{fig:ka}), forming a penny-shaped crack. The fracture dynamics depend on the material properties, such as the Young's modulus $E$, Poisson's ratio $\nu$, and toughness $K_{IC}$, and the injection parameters, i.e. the constant flow rate $Q$ and liquid viscosity $\mu$. As the injection stops, the fracture reaches its final configuration and volume $V_0$. 

 This study addresses the injection of an immiscible liquid in a pre-formed penny-shaped crack. The fluid is injected through the same point source at the centre of the penny-shaped crack. The displaced fluid fills an outer annular region of the fracture (see figure \ref{fig:ka}). In what follows, we use the subscript ``in'' to refer to the injected liquid and ``o'' to the displaced fluid. The surface tension of the interface is noted $\gamma$ and the contact angle with the solid $\theta$. Over the timescale of an experiment, typically a few minutes, the solid is not porous to the liquids and the volume of the fracture is equal to the total volume of fluid injected.

 We make assumptions regarding fluid flow and fracture propagation to model the system dynamics. As the injection rates are low, we assume that the fracture propagates at equilibrium and the liquid and fracture fronts coincide at all times, with no fluid lag. The fluid injection results in linear elastic deformation of the surrounding material. 

 To describe the crack aperture $w(r,t)$, radius $R(t)$ and pressure $p(r,t)$, as well as the position of the interface $R_I$, we need to solve the coupled equations that describe (a) the viscous flow of the two fluid phases in the time-dependent fracture, (b) the elastic deformation of the solid material or fracture walls, (c) the stress intensity factor at the tip of the fracture, and (d) the volume conservation. These sets of equations are coupled by the net pressure in the fracture. The fluid domain is divided into two regions. The outer region is composed of the displaced fluid and bound by the liquid-liquid interface at $r= R_I$ and the crack tip at $r= R$. The injected fluid fills the inner region of the crack from the injection point to the fluid-fluid interface at $r= R_I$. Both regions are axisymmetric as shown in figure \ref{fig:ka}.

\subsection{Fluid flow in the crack}
\subsubsection{Lubrication theory}
 The low Reynolds number flow in the elongated fracture allows to simplify the Navier Stokes equation and use lubrication theory. The fluid is injected in a pre-formed crack whose aspect ratio is small:
 \begin{equation}
  w << R.
 \end{equation}
 The Reynolds number of the flow through the crack is defined for the injected fluid as
 \begin{equation}
 Re = \epsilon  \frac{\rho \, U \, w}{\mu} = \frac{\rho_{in} Q_{in} w(r=0)}{2 \pi \mu_{in} R^2} \leq 1,
 \end{equation}
 where $\epsilon$ is the aspect ratio of the crack. As a result the flow of both fluids can be modeled with the lubrication theory, similarly to the single-phase flows in a penny-shaped fracture \citep{Savitski2002, Detournay2004, Detournay2014, Lai2015, OKeeffe2018a}.
 For the 2-fluid system, the lubrication equations write:
\begin{equation}\label{eq:8}
\frac{\partial w(r, t)}{\partial t}=\frac{1}{12 \mu_{in} } \frac{1}{r} \frac{\partial}{\partial r}\left(r w^{3}(r, t) \frac{\partial p}{\partial r}\right) \; \text {for} \; 0\leq r \leq R_I
\end{equation}
\begin{equation}\label{eq:9}
\frac{\partial w(r, t)}{\partial t}=\frac{1}{12 \mu_{0}} \frac{1}{r}  \frac{\partial}{\partial r}\left(r w^{3}(r, t) \frac{\partial p}{\partial r}\right) \; \text {for} \; R_I\leq r \leq R
\end{equation}

\subsubsection{Liquid interface}
The interface between the two fluids moves outward during the injection and is described by the dynamics boundary condition:
\begin{equation}\label{eq:3}
\boldsymbol{n} \cdot \left(-p^{-} \boldsymbol{I} +\mu_{in}\left(\nabla \boldsymbol{u^{-}}+(\nabla \boldsymbol{u^{-}})^{\mathrm{T}}\right)\right) \cdot \boldsymbol{n}+\gamma \kappa =\boldsymbol{n} \cdot \left(-p^{+}\boldsymbol{I}+\mu_{0}\left(\nabla \boldsymbol{u^{+}}+(\nabla \boldsymbol{u^{+}})^{\mathrm{T}}\right)\right) \cdot \boldsymbol{n}
\end{equation}
where $\bf{n} = \mathbf{e_r}$ is the vector normal to the interface and $\kappa$ the sum of the principal curvatures of the interface. The $+$ and $-$ exponents indicate that the value of the variable is determined at $r = R_I+ \epsilon$ and $r = R_I - \epsilon$ respectively with $\epsilon << R_I$. We note $\boldsymbol{{\tau}} = \mu \left( \nabla \boldsymbol{u}+(\nabla \boldsymbol{u})^{\mathrm{T}} \right)$. We assume that the fluids are perfectly wetting the gel and neglect the thin film deposited by the outer fluid. The normal stress balance writes 
\begin{eqnarray}\label{eq:11}
  p^{-} -  p^{+} &=& \sigma_I  = \gamma \left(\frac{1}{R_I} + \frac{2}{w_I}\right) -  {\bf{n}} \cdot \left(\boldsymbol{\tau^+}-\boldsymbol{\tau^-}\right)\cdot {\bf{n}}  \\
  &=& \gamma \left(\frac{1}{R_I} + \frac{2}{w_I}\right) - 2 \mu_{0} \frac{\partial u}{\partial r}|_{r= R_I^{+}} - 2 \mu_{in} \frac{\partial u}{\partial r}|_{r= R_I^{-}},
\end{eqnarray}
where $w_I$ is the width of the fracture at the interface. The expression can be further simplified as $R_I>w_{I}$ and the viscous normal stresses are negligible. Indeed the capillary number of the invading fluid is
\begin{equation}
Ca_{in} = \frac{\mu_{in} \, U }{\gamma} = \frac{\mu_{in} Q_{in}}{2 \gamma \pi R \, w(r=0)} << 1.
\end{equation}
The pressure change across the interface is:
\begin{equation}\label{eq:50}
  p^{-} -  p^{+} \approx  \frac{2\gamma}{w_I}.
\end{equation}
As the fluid-fluid interface moves, we can write the following kinematic condition using Reynolds equations:
\begin{equation}\label{eq:10}
q^- = q^+\\
 = -\frac{w_I^3}{12\mu_{in}} \frac{\partial p}{\partial r}|_{r= R_I^{-}} = -\frac{w_I^3}{12\mu_{0}} \frac{\partial p}{\partial r}|_{r= R_I^{+}}.
\end{equation}

\subsubsection{Volume conservation}
Finally, through volume conservation, the volume of the crack is equal to the volume injected. We note $V_0$ the volume of the pre-fracture which is equal to the volume of the outer fluid. The injection begins at $t=0$:  
\begin{equation}\label{eq:4}
V_0 + Q_{in}\,t= 2 \pi \int_0^{R}r w(r,t) dr,
\end{equation}
and
\begin{equation}\label{eq:5}
Q_{in}\,t= 2 \pi \int_0^{R_I}r w(r,t) dr.
\end{equation}

\subsection{Fracture equations}
\subsubsection{Linear elasticity}
{Similarly to the single-fluid fracture \citep{Savitski2002}, the linear elasticity equation  writes }
\begin{equation}\label{eq:17}
w(r, t)=\frac{8 \left(1-\nu^2\right) R}{\pi E} \int_{r / R}^{1} \frac{\xi}{\sqrt{\xi^{2}-(r / R)^{2}}} \int_{0}^{1} \frac{x p(x \xi R, t)}{\sqrt{1-x^{2}}} \mathrm{~d} x \mathrm{~d} \xi.
\end{equation}

\subsubsection{Fracture propagation}
In a small region at the tip of the crack, the material undergoes plastic deformation. The tensile fracture propagates when the mode I stress intensity factor $K_I$ reaches a critical value called the toughness of the material $K_{IC} = \sqrt{2 \, E^{'} \, \gamma_S}$, where $\gamma_S$ is the fracture surface energy of the solid material \citep{Kanninen1985} and $E' = E/\left(1-\nu^2\right)$. For a penny-shaped crack, the stress intensity factor near the tip is defined as \citep{Rice1968}:
\begin{eqnarray}\label{eq:17b}
K_{\mathrm{I}}&=&\frac{2}{\sqrt{\pi R}} \int_{0}^{R(t)} \frac{p(r, t)}{\sqrt{R^{2}-r^{2}}} r \mathrm{d} r.
\end{eqnarray}

\subsection{Boundary conditions at the fracture tip and injection point}
At the tip of the crack, the width $w(R)$ goes to zero:
\begin{equation}\label{eq:18}
w=0, \quad r=R(t),
\end{equation}
and the flow rate also goes to zero:
\begin{equation}\label{eq:19}
w^{3}(r, t) \frac{\partial p(r, t)}{\partial r}=0, \quad r=R(t).
\end{equation}

At the point source, the local flow rate is equal to the injected flow rate
\begin{equation}\label{eq:20}
2 \pi \lim _{r \rightarrow 0} r \, q(r, t)=Q_{in}.
\end{equation}

\section{Scaling}
The equations are non-dimensionalized by identifying the characteristic scales in both phases as listed in table \ref{tab:def}.

The characteristic radius and aperture of the fracture are $R$ and $W_0$ respectively, with $R$ the radius of the fracture and $W_0$ the maximum aperture of the fracture at $r=0$. The position of the interface is $R_I$. {The characteristic pressure in both the fluids is taken to be $P_{0}$}. We define effective material parameters ${\mu'}$ $E'$, and $K'$ as proposed by \citet{Savitski2002}:
\begin{equation}\label{eq:21}
\mu_{in}' = 12\mu_{in}, \;
\mu_{0}' = 12\mu_{0}, \;
K' = 4\left(\frac{2}{\pi}\right)^{\frac{1}{2}}K_{IC}, \, \mbox{and}
\; E' = \frac{E}{1-\nu^2}.
\end{equation}
To compare the viscosity of the two liquids, we introduce the parameter $M = \mu_{in}'/\mu_{0}'$. {When the two fluids are present in the fracture, we used a weighted average to define the resulting viscosity $\mu_{e}$ and the corresponding effective value $\mu'_{e} = 12 \mu_{e}$.}

\begin{table}
\begin{center}
 \begin{minipage}{6cm}
\begin{tabular}{lll}
Parameter & Inner region & Outer region \\
Radius      & $r = R_I \hat{r}$     & $r = R \tilde{r}$    \\
Width        & $w= W_0 \hat{w}$     & $w = W_0 \hat{w}$       \\
Pressure      & {$p= P_{0} \hat{p}$}    & {$p= P_{0} \hat{p}$}        \\
\end{tabular}
\caption{\label{tab:def} Rescaled parameters}
 \end{minipage}
 \end{center}
\end{table}

Using the characteristic parameters summarized in table \ref{tab:def} and the effective material properties, we obtain the following set of equations to describe the displacement flow and fracture propagation:

\begin{enumerate}
 \item Lubrication theory (from equations \ref{eq:8}-\ref{eq:9})
 \begin{equation}\label{eq:22}
{\frac{ \mu'_e \, R^2}{W_0^2 \, P_{0}} \; \frac{\partial \hat{w}}{\partial t}=  \frac{1}{\hat{r}} \; \frac{\partial}{\partial \hat{r}}\left(\hat{r} \hat{w}^{3} \frac{\partial \hat{p}}{\partial \hat{r}}\right)} 
\end{equation}

\smallskip
  \item Linear elasticity (from equation \ref{eq:17}) 
    \begin{equation}\label{eq:24}
{\hat{w}  =   \frac{8 R}{ \pi E' W_0} \left[ P_0\int_{0}^{1} \int_{0}^{1} \frac{x \hat{p}(x \xi R,t)}{\sqrt{1-x^2}} dx d\xi   \right]}
\end{equation}

\smallskip
  \item Fracture propagation (from equation \ref{eq:17b})
 \begin{equation}\label{eq:26}
{\frac{K'}{P_{0} \sqrt{R}}= \frac{2^{7/2}}{\sqrt{\pi}}\int_0^{1} \frac{\tilde{p} \; \tilde{r}}{\sqrt{1-\tilde{r}^2}} \, d\tilde{r}}  
\end{equation}
  \smallskip
  \item Global mass balance (from equations \ref{eq:4}-\ref{eq:5})
  \begin{equation}\label{eq:27}
\frac{Q_{in}\,t}{2 \pi \, R_I^2 \, W_0}=  \int_0^{1} \hat{r} \hat{w} \, d\hat{r}
\end{equation}
and
\begin{equation}\label{eq:28}
\frac{V_0}{2 \pi \, R^2 \, W_0} = \int_0^{1} \tilde{r} \hat{w} \, d\tilde{r} - \left(\frac{R_I}{R}\right)^2 \int_0^{1} \hat{r} \hat{w} \, d\hat{r}.\\
\end{equation}
\end{enumerate}
\smallskip
 As the fluid is injected, the elastic pressure drives the propagation of the crack, which is resisted by the viscous dissipation associated with the motion of the injected and displaced fluids and the fracture toughness. We first assume that the interfacial pressure is negligible compared to the viscous and toughness-related stresses. We then assume that one of the resisting stresses controls the propagation and balances the elastic stress. Studies on single fluid injection have validated this approach, with experimental evidence of the two asymptotic regimes. If the viscous stresses in the fluids are larger than the fracture-opening stress, the fracture propagation is said to be in the viscous regime. Alternatively, the propagation is in the toughness regime. For both regimes, we can derive scaling arguments from the dimensionless groups in equations \ref{eq:22}-\ref{eq:28}. 

\subsection{Toughness regime}
For the toughness scaling, we set the non-dimensional groups in equations \ref{eq:24}-\ref{eq:28} equal to one. Indeed the viscous stresses are negligible and the crack opening is limited by the toughness of the material. The scaling relations are:
  \begin{equation}\label{eq:30}
    \frac{W_0E'}{{P_0}R} = 1
\end{equation}
\begin{equation}\label{eq:31}
    \frac{K'}{{P_0} \sqrt{R}}  = 1
\end{equation}
\begin{equation}\label{eq:32}
 \frac{Q_{in}t}{R_I^2 W_0} = 1
\end{equation}
\begin{equation}\label{eq:33}
 \frac{V_0}{ W_0} = R^{2} - R_I^{2}.
\end{equation}
We define the characteristic time scale $T = V_0/Q_{in}$ and the corresponding dimensionless time $\tilde{t} = t/T$. By combining those groups, we obtain the toughness-dominated scaling of the fracture properties:

\begin{equation}\label{eq:34}
    W_0 = \left(\frac{K'}{E'}\right)^{4/5} \, V_0^{1/5} \left(1+\tilde{t}\,\right)^{1/5},  
\end{equation}
\begin{equation}\label{eq:35}
  R  =  \left(\frac{E' \, V_0}{K'}\right)^{2/5} \, \left(1+\tilde{t}\,\right)^{2/5},   
\end{equation}
\begin{equation}\label{eq:36}
    R_I = \left(\frac{E' \, V_0}{K'}\right)^{2/5} \, \tilde{t}^{\,1/2} \, \left(1+\tilde{t}\,\right)^{-1/10},
\end{equation}
and
\begin{equation}\label{eq:37}
    {P_0} = K' \left(\frac{K'}{E' \, V_0}\right)^{1/5} \, \left(1+\tilde{t}\,\right)^{-1/5}.
\end{equation}

\begin{table}
\centering
\begin{tabular}{cll}
Fracture  & \multicolumn{1}{c}{Viscous regime} & \multicolumn{1}{c}{Toughness regime} \\
\multicolumn{1}{l}{} &                                    &                                      \\
Radius &
  $ { \hspace{2.5mm}   R \approx  \left(\frac{E'}{\mu'_{e}Q_{in}}\right)^{1/9}  V_0^{4/9} \left(1+\tilde{t}\right)^{1/3} \left(\alpha+\tilde{t}\right)^{1/9}} $ &
  $ \hspace{0.5mm} R \; \approx  \left(\frac{E' \, V_0}{K'}\right)^{2/5} \, \left(1+\tilde{t}\,\right)^{2/5}$ \\
\multicolumn{1}{l}{} &                                    &                                      \\
Interface &
  $ \hspace{0.75mm}   {R_I \approx \left(\frac{E'}{\mu'_{e}Q_{in}}\right)^{1/9}  V_0^{4/9} \sqrt{\tilde{t}} \left(1+\tilde{t}\right)^{-1/6}  \left(\alpha+\tilde{t}\right)^{1/9}}$ &
  $R_I \approx \left(\frac{E' \, V_0}{K'}\right)^{2/5} \, \tilde{t}^{\,1/2} \left(1+\tilde{t}\,\right)^{-1/10}$ \\
\multicolumn{1}{l}{} &                                    &                                      \\
Aperture &
  $ { \hspace{0 mm}  W_0 \approx \left(\frac{\mu'_{e}Q_{in}}{E'}\right)^{2/9}V_0^{1/9} \left(1+\tilde{t}\right)^{1/3} \left(\alpha+\tilde{t}\right)^{-2/9}}$ &
  $W_0 \approx \left(\frac{K'}{E'}\right)^{4/5} \, V_0^{1/5} \left(1+\tilde{t}\,\right)^{1/5}$
\end{tabular}
\caption{Scaling relations for a penny-shaped fracture driven by a displacement flow. The time evolution of the geometrical properties of the fracture depends on the dominant resisting stress, which can be viscous or toughness-related.}
\label{tab:my-table}
\end{table}

\subsection{Viscous regime}
For the viscous scaling, we set the non-dimensional groups in equations 3.2-3.3 and \ref{eq:27}-\ref{eq:28} equal to one. Here the propagation of the fluids follows the lubrication equation and the viscous dissipation limits the propagation of the fracture:
\begin{equation}\label{eq:38}
   { \frac{W_0E'}{P_{0}R}= 1}
\end{equation}

\begin{equation}\label{eq:39}
{  \frac{ \mu'_{e} \, R^2}{W_0^2 \, P_{0}\, \left(t+V_0/Q_{0}\right) } = 1}
\end{equation}
\begin{equation}\label{eq:41}
{ \frac{Q_{in}t}{ R_I^2 W_0} = 1}
\end{equation}
\begin{equation}\label{eq:42}
{ \frac{V_0}{ W_0} = R^{2} - R_I^{2} }
\end{equation}
\smallskip
{We define the effective viscosity of the volume of fluid in the fracture as a weighted average. The effective viscosity depends on the viscosities of the two fluids in the fracture and their relative volumes at time $\tilde{t}$: 
\begin{equation}\label{eq:48}
   \mu'_e = \frac{\mu'_{0}+\mu'_{in}\tilde{t}}{1 + \tilde{t}}
\end{equation}}

We obtain the viscous-dominated scaling of the variables:
\begin{equation}\label{eq:43}
  {  W_0 = \left(\frac{\mu'_{e}Q_{in}}{E'}\right)^{2/9}  V_0^{1/9} \left(1+\tilde{t}\right)^{1/3} \: \left(\alpha+\tilde{t}\right)^{-2/9} ,  }
\end{equation}

\begin{equation}\label{eq:44}
 {   R =  \left(\frac{E'}{\mu'_{e}Q_{in}}\right)^{1/9}  V_0^{4/9} \; \left(1+\tilde{t}\right)^{1/3} \: \left(\alpha+\tilde{t}\right)^{1/9}  , } 
\end{equation}
\begin{equation}\label{eq:45}
  {  R_I =  \left(\frac{E'}{\mu'_{e}Q_{in}}\right)^{1/9}  V_0^{4/9} \; \sqrt{\tilde{t}} \: \left(1+\tilde{t}\right)^{-1/6} \: \left(\alpha+\tilde{t}\right)^{1/9} ,}
\end{equation}
\begin{equation}\label{eq:46}
  {  P_{0} = \left(\frac{\mu'_{e}Q_{in}E^{\prime 2}}{V_0}\right)^{1/3}\left(\alpha + \tilde{t}\right)^{-1/3}}
\end{equation}

with $\alpha = Q_{in}/{Q_{0}}$. The results are summarized in table \ref{tab:my-table}.

\subsection{Discussion}
We consider the single-fluid limit of the two asymptotic regimes. Indeed for a pre-fracture volume equal to zero, $V_0=0$, or for large values of the injection time $t>>1$, the expressions derived above should be equal to those previously obtained for single fluid injection. In the toughness regime, we recover the single-fluid scaling relations \citep{Savitski2002,Lai2016a,OKeeffe2018a}:
\begin{equation}\label{eq:53b}
    W_0 = \left(\frac{K'}{E'}\right)^{4/5} \, \left(Q_{in}\,t\right)^{1/5},  
\end{equation}
\begin{equation}\label{eq:54}
    R =  \left(\frac{E'}{K'}\right)^{2/5} \, \left(Q_{in}\,t\right)^{2/5},     
\end{equation}
and
\begin{equation}\label{eq:55}
    P_0 = K' \left(\frac{K'}{E' \, Q_{in} \,t}\right)^{1/5}.
\end{equation}
In the viscous regime, we also recover the scaling relations for a single fluid injection:
\begin{equation}\label{eq:56}
    W_0 = \left(\frac{\mu'_{in}}{E'}\right)^{2/9}  Q_{in}^{1/3} \, t^{1/9},  
\end{equation}
\begin{equation}\label{eq:57}
    R =  \left(\frac{E'}{\mu'_{in}}\right)^{1/9}  Q_{in}^{1/3} \, t^{4/9}  ,  
\end{equation}
and
\begin{equation}\label{eq:58}
    P_0 = \left(\frac{\mu'_{in}E^{\prime \, 2}}{t}\right)^{1/3}.
\end{equation}

{The fluid and matrix properties determine whether the fracture propagation is in the viscous or toughness regime. Past studies in single fluid injection have defined criteria to predict the propagation regime. The regime is defined by the largest of the two stresses that oppose the elastic stress: the toughness-related  stress $\Delta P_m \approx \frac{K'}{\sqrt{R}}$ and the viscous stress $\Delta P_v = \frac{\mu'Q}{W_0^3}$. In the viscous regime, we can use the scaling relations for $R$ and $W_0$ to estimate the ratio:
\begin{equation}\label{eq:58b}
  \left( \frac{\Delta P_m}{\Delta P_v}\right)_v = \left(\frac{K^{\prime\,9} \, t}{E^{\prime 13/2} \, Q^{3/2} \,  \mu^{\prime 5/2}}\right)^{1/9} = \left(\frac{t}{t_{mk}}\right)^{1/9}= \kappa.
\end{equation}
where
\begin{equation}\label{eq:58c}
  t_{mk} = \left(\frac{E^{\prime 13/2} \, Q^{3/2} \, \mu^{\prime 5/2}}{K^{\prime\,9} }\right)
\end{equation}
is the characteristic timescale of the system and $\kappa$ is the dimensionless toughness. Similarly, we can define the relative magnitude of both stresses in the toughness regime using the corresponding scaling relations:
\begin{equation}\label{eq:58c}
  \left( \frac{\Delta P_m}{\Delta P_v}\right)_m =  \left(\frac{t}{t_{mk}}\right)^{2/5}= \kappa^{18/5}.
\end{equation}
The viscous-dominated propagation is therefore associated with small values of $\kappa$, i.e. $\kappa \; \mbox{or} \; t/t_{mk} \lesssim 1$ and the toughness-dominated dynamics for $\kappa \; \mbox{or} \; t/t_{mk} >>1$.\\
 Similarly, we can define the propagation regime of the fracture formed by a displacement flow. The toughness related stress remains $\Delta P_m \approx K'/\sqrt{R}$, while the viscous stress becomes \\ $\Delta P_v =\mu_e'Q_{in}  {R^2}\left({R_I^2 \, W_0^3+ \alpha V_0 \, W_0^2 }\right)^{-1}$ as defined in equations \ref{eq:31} and \ref{eq:39}.\\
In the viscous regime, we substitute $R$, $R_I$, and $W_0$ by the scaling relations defined in table 3: 
\begin{equation}\label{eq:58d}
   \left(\frac{\Delta P_m}{\Delta P_v}\right)_v = 
   \frac{K' V_{0}^{1/9}(\alpha+\tilde{t})^{5/18}}{E^{\prime 13/18} \mu_{e}^{\prime 5/18}Q_{in}^{5/18}(1 +\tilde{t})^{1/6}}
 \end{equation}
Similarly, in the toughness regime, we estimate the ratio to be
\begin{equation}\label{eq:58e}
   \left( \frac{\Delta P_m}{\Delta P_v}\right)_m =  \left(\frac{\Delta P_m}{\Delta P_v}\right)_v^{18/5}
 \end{equation}
The ratio of the viscous and toughness-related stresses is a function of time, that increases as $\tilde{t}^{1/9}$ for large values of $\tilde{t}$. In consequence, the propagation becomes toughness-controlled for long time or large-volume injections. This result is consistent with what is known for a single fluid injection. Contrary to the single fluid criterion however, there is no explicit solution for the threshold injection time in the case of displacement flow. We can determine the propagation regime at time $\tilde{t}$ by comparing the pressure ratio with 1.}

\section{Experiments}
Laboratory-scale experiments commonly use hydrogels as rock analogues to study hydraulic fracturing in brittle elastic materials \citep{OKeeffe2017, Baumberger2020}. In particular, gelatin is a clear gel whose elasticity can easily be tuned by varying the volume fraction of gelatin powder in water \citep{Giuseppe2009, Kavanagh2013, Lai2015}. Because gelatin expands as it sets in the container, the material is spontaneously compressed, which furthers the analogy with soft rocks for fracture studies.

\begin{sidewaystable}
\centering
\begin{tabular}{clclccclccc}
\hline
\multicolumn{1}{l}{} &
  &
  \multicolumn{1}{c}{Gel} &
  \multicolumn{1}{r}{} &
  \multicolumn{3}{c}{Pre-fracture - oil-filled} &
  \multicolumn{1}{r}{} &
  \multicolumn{3}{c}{Injection} \\
\multicolumn{1}{l}{Exp.} &
  &
  $E$ (kPa) &
  &
  $ \mu_0$ (Pa.s) &
  $Q_0$ (ml.min$^{-1}$) &
  $V_0$ (ml) &
  &
  Fluid &
  \multicolumn{1}{l}{$\mu_{in}$ (Pa.s)} &
  \multicolumn{1}{l}{$Q_{in}$ (ml.min$^{-1}$)} \\
  \hline
\multicolumn{1}{l}{} &
  &
  \multicolumn{1}{l}{} &
  &
  \multicolumn{1}{l}{} &
  \multicolumn{1}{l}{} &
  \multicolumn{1}{l}{} &
  &
  \multicolumn{1}{l}{} &
  \multicolumn{1}{l}{} &
  \multicolumn{1}{l}{} \\
1  & $\color{yellow1} {\mathlarger{\mathlarger{\bullet}}} \hspace{-2.5 mm} \color{black} {\mathlarger{\mathlarger{\circ}}}$      & 30  &  & $0.01$ & 0.3  & 5.6 &  & Water & 0.001 & 0.5  \\
 2  & {\color{orange1}\rotatebox[origin=c]{90}{$\mathlarger{\mathlarger{\blacktriangle}}$}} \hspace{-2.5 mm} \raisebox{-1.3pt}{{\begin{rotate}{90} $\bigtriangleup$ \end{rotate}}} & 30  &  & 0.01 & 0.3  & 6.2 &  & Water & 0.001 & 1 \\
 3  &   $\color{black} {\mathlarger{\mathlarger{\blacktriangle}}}   \hspace{-5.1 mm} \color{black}$ {\raisebox{-0.5pt}{{$\bigtriangleup$}}}           & 30  &  & $0.01$ & 0.15 & 5.1 &  & Water & 0.001 & 0.15 \\
 4  &  {\color{purple1}{\rotatebox[origin=c]{-90}{$\mathlarger{\mathlarger{\blacktriangle}}$}}} \hspace{-4.45mm}{\raisebox{9.3pt}{{\begin{rotate}{-90} $\bigtriangleup$ \end{rotate}}}} & 30  &  & $0.01$ & 0.15 & 6.3 &  & Water & 0.001 & 1    \\
 5  &  {\color{green1}{\rotatebox[origin=c]{180}{$\mathlarger{\mathlarger{\blacktriangle}}$}}} \hspace{-1.4 mm}{\raisebox{8.5pt}{{\begin{rotate}{180} $\bigtriangleup$ \end{rotate}}}}   & 15  &  & $0.02$ & 0.1  & 7.1 &  & Water & 0.001 & 0.25 \\
 6  &  $\color{blue1} {\hspace{1 pt}{\blacklozenge}} \color{black} {\hspace{-8  pt}{\lozenge}} $  & 30  &  & $0.01$ & 0.3  & 3.9 &  & Water & 0.001 & 1.8  \\
7  & $\color{crimson1} \times$   & 30  &  & $0.01$ & 0.15 & 4   &  & Water & 0.001 & 1.8  \\
 8  & $ \blacksquare $    & 30  &  & $0.01$ & 0.5  & 6.2 &  & Water & 0.001 & 3.6  \\
 9  &  $\color{brown1} +$   & 30  &  & $0.01$ & 0.15 & 2.9 &  & Water & 0.001 & 0.54 \\
 10 &  {\color{lavender1}{\rotatebox[origin=c]{-90}{$\mathlarger{\mathlarger{\blacktriangle}}$}}} \hspace{-4.45 mm}{\raisebox{9pt}{{\begin{rotate}{-90} $\bigtriangleup$ \end{rotate}}}}& 15  &  & $0.01$ & 0.25 & 6.1 &  & Water & 0.001 & 0.5  \\
 11 &    $\color{darkblue1}$ \hspace{-2 pt}{{\raisebox{0.5pt}{$\times {\hspace{-3.35mm} +}$}}}  & 88  &  & $10.26$   & 10   & 10 &  & Glycerol & $1.2$ & 10   \\
 12 & $\color{peach1} \blacksquare \hspace{-3.2mm}\color{black} \square $ & 88  &  & $10.26$   & 10   & 15 &  & Glycerol/water & $0.3$ & 10   \\
 13 & $\color{black} {\mathlarger{\mathlarger{\bullet}}} \hspace{-2.4 mm} \color{black} {\mathlarger{\mathlarger{\circ}}}$    & 88  &  & $10.26$   & 10   & 15  &  & Glycerol & $1.2$ & 10   \\
 14 &    $\color{teal1} {\mathlarger{\mathlarger{\blacktriangle}}}   \hspace{-5.1 mm} \color{black}$ {\raisebox{-0.25pt}{{$\bigtriangleup$}}}    & 88 & & $10.26$   & 10  & 5 &  & Glycerol & $1.2$ & 10   \\
15 & $\color{red1} {\mathlarger{\mathlarger{\star}}}$  & 88  &  & $10.26$   & 10 & 6.5   &  & Syrup & $8.6$ & 10  \\
 16  & \color{black}  {\rotatebox[origin=c]{90}{$\mathlarger{\mathlarger{\blacktriangle}}$}}  \hspace{-2.5 mm} \raisebox{-1.5pt}{{\begin{rotate}{90} $\bigtriangleup$ \end{rotate}}} & 88  & & $10.26$ & 10 & 8 &  & Syrup & $8.6$  &  10 \\
\end{tabular}
\caption{List of experiments: symbols and parameters.}
\label{tab:my-exp}
\end{sidewaystable}

\subsection{Material preparation and characterisation}
 The properties of the gelatin and the injected fluids control the fracture dynamics and are, therefore, systematically characterized. The gelatin is prepared by heating ultra-pure water to $60^{o}$C. While stirring the heated water, we slowly add gelatin powder (Gelatin type A; Sigma-Aldrich, USA) at a mass fraction of $10$ to $30\,\%$ to vary Young's modulus of the resulting gel. The gelatin then cools down to room temperature and sets over 24 hours prior to testing or fracturing. The Young's modulus of gelatin is measured with a custom-built displacement-controlled load frame. The sample is compressed by a ballscrew stage whose speed is set by a stepper motor Parker Compumotor OS22B controlled by a controller Parker Compumotor ZL6104. A load cell,  Eaton 3108-10 (10 lb. capacity) measures the force generated by the compressed sample. We record force-displacement values for cylindrical samples of gelatin of diameter and height equal to 1 in and compute the stress-strain curves of the material. At small strain, all gelatin samples exhibit a linear elastic response to the compression. As listed in table \ref{tab:my-exp}, Young's modulus ranges between $15$ to $116$ kPa with a measurement error of $\pm 10 \, \%$. The fracture energy and Poisson's ratio of the gelatin are assumed constant in our experiments with $\gamma_S \approx 1$ J.m$^{-2}$ and $\nu \approx 0.5$ \citep{Menand2002}.

 To study displacement flows in fluid-filled fractures, we use immiscible Newtonian liquids. Silicone oils of different viscosity are used to form the pre-fracture. An aqueous solution composed of water, glycerol or corn syrup is then injected. The silicone oil is dispalced outward, further expanding the fracture in the gelatin. We measure the viscosity of the fluids and the surface tension at the oil/water and oil/syrup interfaces. Viscosity measurements are conducted using an MCR 92 Anton Parr rheometer with a parallel plate measuring system at $20^o$C. The values obtained have a measurement error of $\pm 1 \, \%$ and are listed in table \ref{tab:my-exp}. Surface tension measurements are conducted using the pendant drop method with an Attension Theta Flex tensiometer: the surface tension between silicone oil and water is $\gamma_{o/w} \approx 35 \pm 2$ mN.m$^{-1}$ and between silicone oil and syrup is $\gamma_{o/s} \approx 50 \pm 2$ mN.m$^{-1}$. {The fluid properties are selected to study the propagation of the pre-fracture and fracture in a single regime of during the experiment. For example, the values of $\kappa$ at the end of the injection forming the pre-fracture, for experiments 1 through 10 varies between $ 35 \leq \kappa \leq 55 $. Those experiments are expected to be in the toughness dominated regime. Similarly, for experiments 11 through 16, $1.7 \leq \kappa \leq 1.9$ at the end of the formation of the pre-fracture: these experiments target the viscous-dominated propagation. These values are consistent with previous work on the formation of fluid-driven fractures in a bloc of gelatin. During the formation of the fracture through the displacement flow, we estimate the pressure ratio in the viscous regime $\left( \frac{\Delta P_m}{\Delta P_v}\right)_v$ at the end of the injection. The ratio varies between 24 and 58 for experiments 1 through 10, which are therefore expected to be in the toughness regime. The ratio varies between 1.9 and 2.5 for experiments 11 through 16, which are therefore expected to be in the viscous regime. }

\begin{figure}
  \centerline{\includegraphics[width=0.9\linewidth]{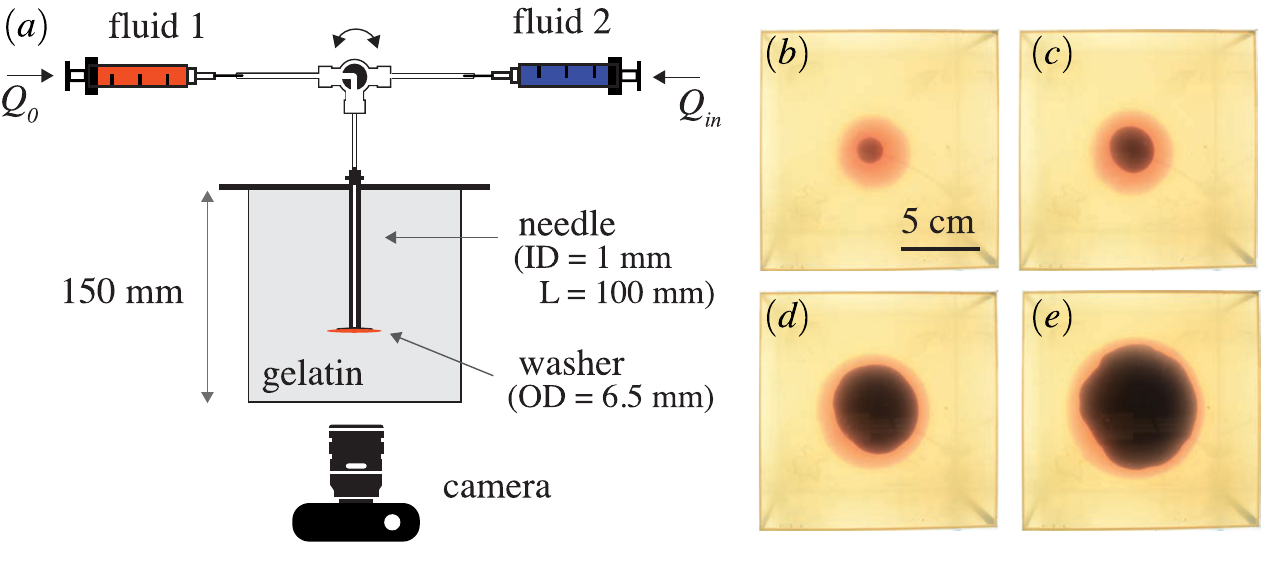}}
  \caption{($a$) Schematic of the experimental set-up. ($b-e$) Time evolution of the fracture formed in experiment 1. The water dyed with blue food colour is injected at $Q_{in} = 0.5$ ml.min$^{-1}$ in a pre-fracture formed with silicone oil dyed with red food colour. The recording starts when the water injection begins and the experimental images are taken at ($b$) $t =50$ s, ($c$) $t = 300$ s, ($d$) $t = 1300$ s  and  at ($e$) $t =2800$ s. The supplementary movies 1 and 2 show the complete time evolution of the pre-fracture and fracture, respectively. The movies are available at [to be added].}
\label{fig:setup}
\end{figure}

\subsection{Set-up}
 The injection experiments are conducted in a large block of gelatin to avoid boundary effects on the propagation of the fracture. The gelatin is set in a cubic clear container (15 cm $\times$ 15 cm $\times$ 15 cm) around a blunt needle as represented in figure \ref{fig:setup}. We use a thin needle (inner diameter, $ID = 1$ mm) for low-viscosity injections and a wide needle for high-viscosity injection ($ID = 2.15$ mm). In our system, the confining stress is minimum in the vertical direction, the fracture, therefore, propagates horizontally, in the direction that opposes the least resistance. To avoid small tilts of the fracture that would compromise the quality of the recording, a plastic washer is placed at the tip of the needle to initiate the fracture in the horizontal plane.
 Two fluids are successively pumped into the gelatin. For each fluid, we use a kdScientific\textsuperscript{\tiny\textregistered} Legato 200 syringe pump to set the injection flow rate at a value between $0.1$ and $20$ ml.min$^{-1}$ with an accuracy of $\pm 0.35\,\%$. Both fluid-filled syringes are connected to the same injection needle using a switch valve. First, a silicone oil of viscosity $\mu_0$ is injected at a constant volumetric flow rate $Q_0$ in the gelatin matrix to form the pre-fracture. The injection stops when a volume $V_0$ of silicone oil has been dispensed. The valve is then switched to inject the aqueous phase of viscosity $\mu_{in}$ into the pre-fracture at a flow rate of $Q_{in}$. The injection stops when the fracture tip is within 2 cm of the container walls to avoid confinement effects \citep{Bunger2013}. The two fluids are dyed to allow visualizing the propagation of the fracture and interface between the two liquids in the clear gelatin: we use a blue water-based food dye for the aqueous phase and a red oil-based food dye for the oil phase as shown in figure \ref{fig:setup}. The propagation is recorded using a Nikon D5300 camera with a Phlox\textsuperscript{\tiny\textregistered} LED panel ensuring uniform backlighting. The images are processed using a custom-made MATLAB code to determine the radius of the fracture $R$ and the position of the interface between the two liquids $R_I$.

 \subsection{Measurement of the fracture width}
 To measure the thickness of the fracture during the propagation, we use a light absorption method \citep{Bunger2006, Bunger2008}. This method consists in selecting a soluble dye and the corresponding optical filter. The filter should transmit light to the camera at the wavelength at which the dye absorbance $A$ is maximum $A_{\lambda}$. The absorbance follows Beer's law:
\begin{equation}
    A_{\lambda} = - \log_{10}\left(\frac{I_{\lambda}}{I_{\lambda,0}}\right)= \epsilon_{\lambda} c \, h
\end{equation}
 where $I_{\lambda,0}$ is the background intensity and $I_{\lambda}$ is the intensity when light passes through a liquid layer of thickness $h$, with a dye concentration $c$. The parameter $\epsilon_{\lambda}$ characterizes the absorbance of the dye at the selected wavelength and is obtained through calibration.
 To measure the thickness of both liquids in the fracture, we use two dyes, one water-soluble and one oil-soluble and record the absorbance using a single optical filter. To get accurate measurements, we select dyes with a large absorbance at the same wavelength. In all experiments, the water-soluble dye is nigrosin (Sigma-Aldrich) at 0.05 g.L$^{-1}$. Nigrosin is a black dye that absorbs at all wavelengths. We use different dyes depending on the viscosity of the silicone oil, because of their solubility limit. We dilute sudan red (Sigma-Aldrich) in the low-viscosity silicone oils, i.e. 10 and 20 mPa.s silicone oils, at 0.05 g.L$^{-1}$, and nile red (Sigma-Aldrich) in  high-viscosity silicone oils, i.e. 10,300 and 30,000 mPa.s silicone oils, at 0.2 g.L$^{-1}$. For experiments with sudan red in the oil phase and nigrosin in the aqueous phase, we use a 520 nm optical filter. When nile red dyes the oil phase and nigrosin the aqueous phase, we use a 632 nm optical filter.

 \begin{figure}
 \includegraphics[width=1\linewidth]{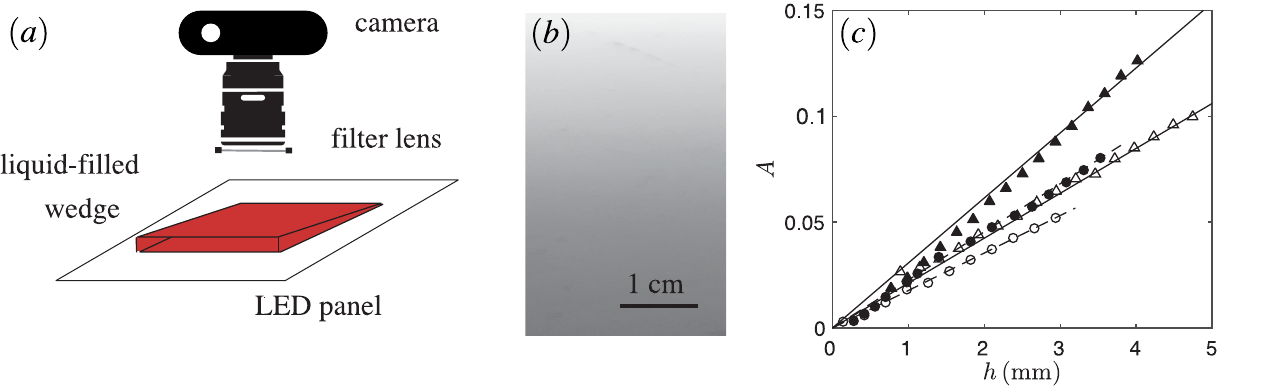}
  \caption{Calibration. ($a$) Schematic of the calibration experiment for low-viscosity fluids. ($b$) The intensity gradient was recorded for a wedge filled with water dyed with nigrosin at 0.05 g.L$^{-1}$ through the 520 nm filter. ($c$) Absorbance measured for nigrosin-dyed water at $\lambda = 520$ nm ($\blacktriangle$), nigrosin-dyed syrup at $\lambda = 632$ nm (${{\mathlarger{\bullet}}}$), sudan red-dyed 10 mPa.s-silicone oil at $\lambda = 520$ nm ($\vartriangle$), and nile red-dyed 10 Pa.s-silicone oil at $\lambda = 632$ nm (${{\mathlarger{\circ}}}$). The solid lines are the best linear fit for each calibration data set.}
 \label{fig:calic}
 \end{figure}

To measure the thickness of the fracture, we first conduct calibration experiments for each dye solution. For the low-viscosity solutions (water, 10 and 20 mPa.s silicone oils), we use a glass wedge with an aperture that increases linearly from 0 to 10 mm as shown in figure \ref{fig:calic}. The wedge is placed on the LED panel and the light intensity is obtained by taking a picture of the wedge with the optical filter mounted on the camera. The background intensity corresponds to an empty wedge. The light intensity of a liquid-filled wedge decreases as the thickness of the aperture increases (see figure \ref{fig:calic}(b)). The grey values are used to determine the absorbance as a function of the liquid thickness, as plotted in figure \ref{fig:calic}(c). Using a linear fit, we get 1/${\epsilon_{\lambda} c}$ = 47.2 mm for sudan red in oil and 1/${\epsilon_{\lambda} c}$ = 32.6 mm for nigrosin in water for $\lambda = 520$ nm.
For the high-viscosity samples, we use rectangular cells that are easier to fill. The cell thickness or height ranges from 0.3 to 3 mm. For each cell, we measure the background intensity of the empty cell and the intensity of the cell filled with the viscous fluid, i.e. syrup, 10,000 mPa.s or 30,000 mPa.s silicone oil. We obtain the absorbance for a set of thickness values as shown in figure \ref{fig:calic}. Using a linear fit, we get 1/${\epsilon_{\lambda} c}$ = 56.5 mm for nile red in oil and 1/${\epsilon_{\lambda} c}$ = 44 mm for nigrosin in water for $\lambda = 632$ nm. The fitting parameters obtained are then used to obtain the width of the fracture using Beer's law:
 \begin{equation}
   h =  \frac{A_{\lambda}}{\epsilon_{\lambda} c} = \frac{- \log_{10}\left(\frac{I_{\lambda}}{I_{\lambda,0}}\right)}{\epsilon_{\lambda} c}
 \end{equation}

 The concentrations of dyes chosen for this study, some of which are limited by solubility, are all sufficiently low for the absorbance to vary linearly with the sample thickness or fracture aperture. The accuracy of the measurements is limited by the noise due to low absorbance values.

 \begin{figure}
\begin{center}
{\includegraphics[width=1\linewidth]{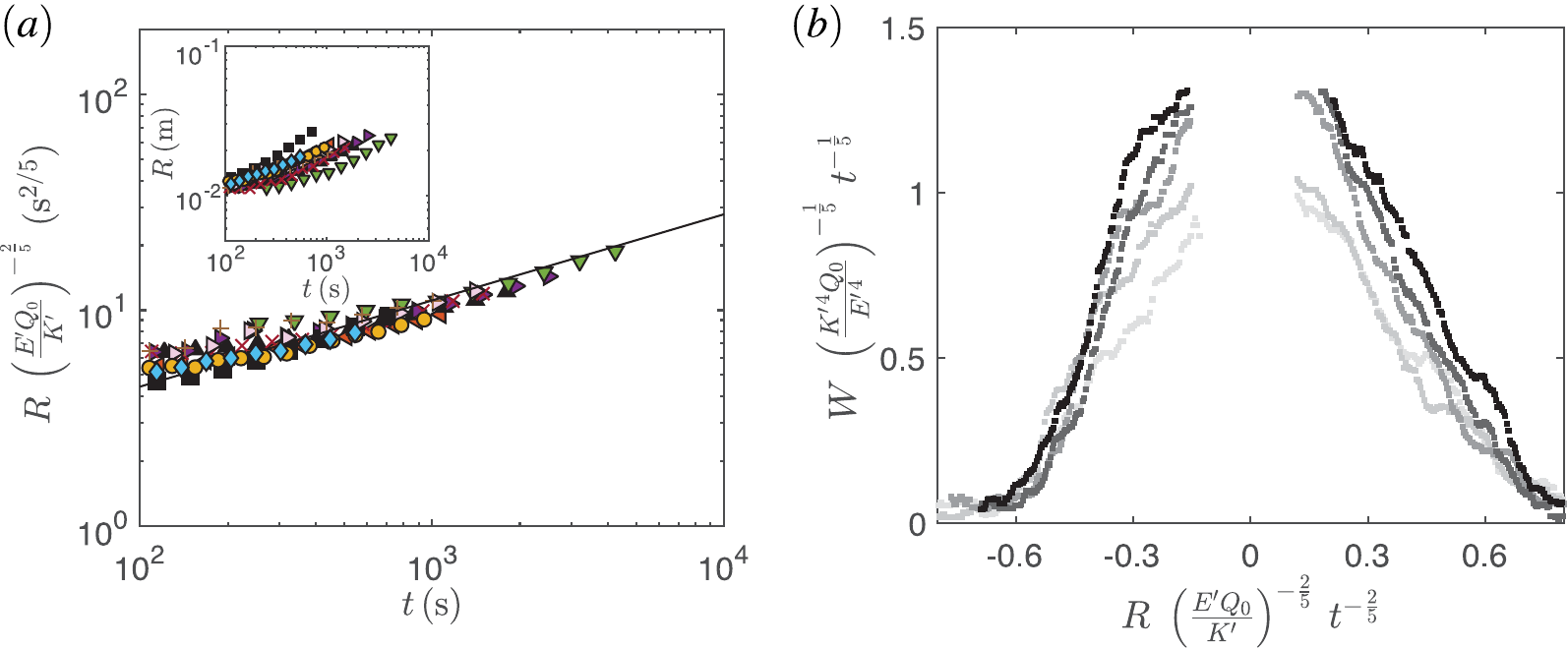}} 
\caption{Dynamics of the pre-fracture for low-viscosity oils. (a) Dependence of rescaled fracture radius on time for experiments 1 to 10 (see table \ref{tab:my-exp} for corresponding injection parameters). The radius is rescaled using equation \ref{eq:54} in the toughness scaling for single fluid injection. The origin for time is set when the oil enters the gelatin. The black curve represents the best linear fit with a slope of 2/5. Inset: dependence of the radius R on time t.  (b) Rescaled fracture thickness profiles based on equations \ref{eq:53b} and \ref{eq:54} at $t = [245,\,370,\,495,\,620,\,745]$ s with time increasing from clear to dark gray. Experimental parameters: $E=30$ kPa, $\mu_0= 10$ mPa.s, $Q_0 = 0.3$ ml.min$^{-1}$.}
\label{fig:SFT}
\end{center}
\end{figure}

\section{Results}
In this section, we present the results of the experiments listed in table \ref{tab:my-exp}. We measure the radius and aperture of the pre-fracture and fracture and compare the data with the scalings obtained for the viscous and toughness regimes. 

The fracture propagates radially upon injection of the fluid. Viscous fingering is not observed in our experiments, as illustrated in figure \ref{fig:setup}(b)-(e). The viscosity ratio $M = \frac{\mu_{in}}{\mu_{out}}$ varies between 0.05 and 1, and the thickness of the crack is of the order of a few millimetres, so the characteristic wavelength of the instability is larger than the perimeter of the injected fluid region. {We estimate the experimental error by conducting an error analysis based on the scaling laws and the measurement error of the various parameters: $10\%$ for the Young's modulus $E$, $10\%$ for the fracture toughness $K$, $0.35\%$ for the flow rate $Q$, $1\%$ for the viscosity $\mu$ and $5\%$ for the volume $V_0$. In the toughness regime, the experimental error on the radius is estimated to be $\sim 10\%$. In the viscous regime, the error is $\sim 4\%$.}

\subsection{Single Fluid Injection}
\begin{figure}
\begin{center}
{\includegraphics[width=1\linewidth]{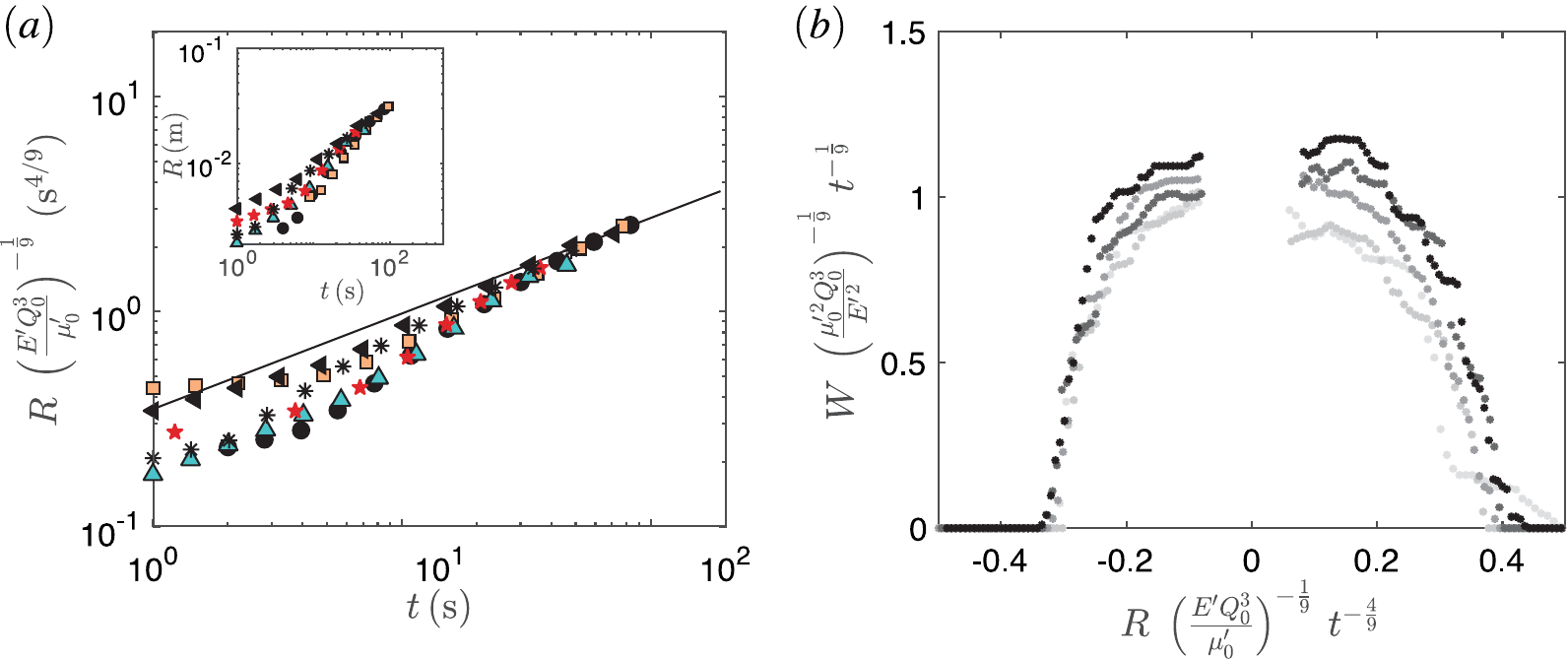}} 
\caption{Dynamics of the pre-fracture for high-viscosity oils. (a) Dependence of rescaled fracture radius on time for experiments 11 to 16 (see table \ref{tab:my-exp} for corresponding injection parameters). The radius is rescaled using equation \ref{eq:57} in the viscous scaling for single fluid injection. The black curve represents the best linear fit with a slope of 4/9. Inset: dependence of the radius R on time t.  (b) Rescaled fracture thickness profiles based on equations \ref{eq:56} and \ref{eq:57} at $t = [9,\,11.5,\,14,\,16.5,\,19]$ s with time increasing from clear to dark gray. Experimental parameters: $E=88$ kPa, $\mu_0= 10$ Pa.s, $Q_0 = 10$ ml.min$^{-1}$.}
\label{fig:SFV}
\end{center}
\end{figure}

We prepare the pre-fracture by injecting silicone oil into the gelatin cube. After an initial pressure build-up, the oil propagates rapidly over the washer at the tip of the needle. A radial fracture then forms around the washer, propagating more slowly with the oil filling the gap between the two gelatin surfaces. The homogeneous properties of the gelatin result in axisymmetric fractures for both low and high-viscosity fluids.
The radius of the fracture is measured during the injection and compared with the scaling for the toughness and viscous-dominated regimes for a single fluid (equations \ref{eq:54} and \ref{eq:57}, respectively). Experiments 1-10 are in the toughness regime as low-viscosity silicone oil is injected in soft gelatin (see table \ref{tab:my-exp}). Experiments 11-16 are in the viscous regime as high-viscosity silicone oil is injected in harder gelatin. In figure \ref{fig:SFT}(a), we plot the results of the experiments in the toughness regime. The radius increases with time and the rescaled radius follows a $t^{2/5}$ power law which is consistent with equation \ref{eq:54}. In the log-log scale, the best fit line with a slope $2/5$ has a prefactor $k = 0.7$ which is in {agreement} with the theoretical prefactor of 0.85 derived by \citet{Savitski2002}. These results are also consistent with previous experimental data in this regime \citep{Lai2016a, OKeeffe2018a}. In figure \ref{fig:SFV}(a), we report the data obtained in the viscous regime. The experimental conditions for those experiments are similar and the results demonstrate the high reproducibility of the experiments \cite{Lecampion2017}. The radius increases as a power law of time. Upon rescaling the radius with the viscous scaling parameter, the data collapse on a line of slope $4/9$ with a prefactor of $k=0.36$. This result differs from the theoretical prefactor of $0.7$, yet it is comparable to the values obtained in previous experimental studies in the viscous regime \citep{Lai2015, OKeeffe2018a}.  Due to the initiation transient and finite size of the container, the power-law fits span about a decade of the log-log plots, which is common for laboratory-scale experiments.

For each regime, we measure the fracture aperture using a dye whose absorbance varies linearly with the aperture. The results presented show the evolution of the fracture cross-section over time for one experiment in the toughness regime (see figure \ref{fig:SFT}(b)) and one in the viscous regime (see figure \ref{fig:SFV}(b)). For both experiments, the curves show the aperture as a function of the radial position at different injection times. Because the needle disturbs the absorbance measurement near the center of the fracture, the aperture is not measured near the needle, i.e. for small values of the radius. Upon integration of the thickness curve recorded when the injection is complete, we obtain the value of $V_0 \pm 0.5$ ml. The aperture-radius curves are rescaled using the scaling for the aperture and radius. In both regimes, the curves collapse on a self-similar profile. The two data sets presented here are representative of the prefacture obtained for all the experiments conducted in this study and are similar to results previously reported \citep{Lai2016a, OKeeffe2018a}. 

\begin{figure}
\centering
{\includegraphics[width=0.9\linewidth]{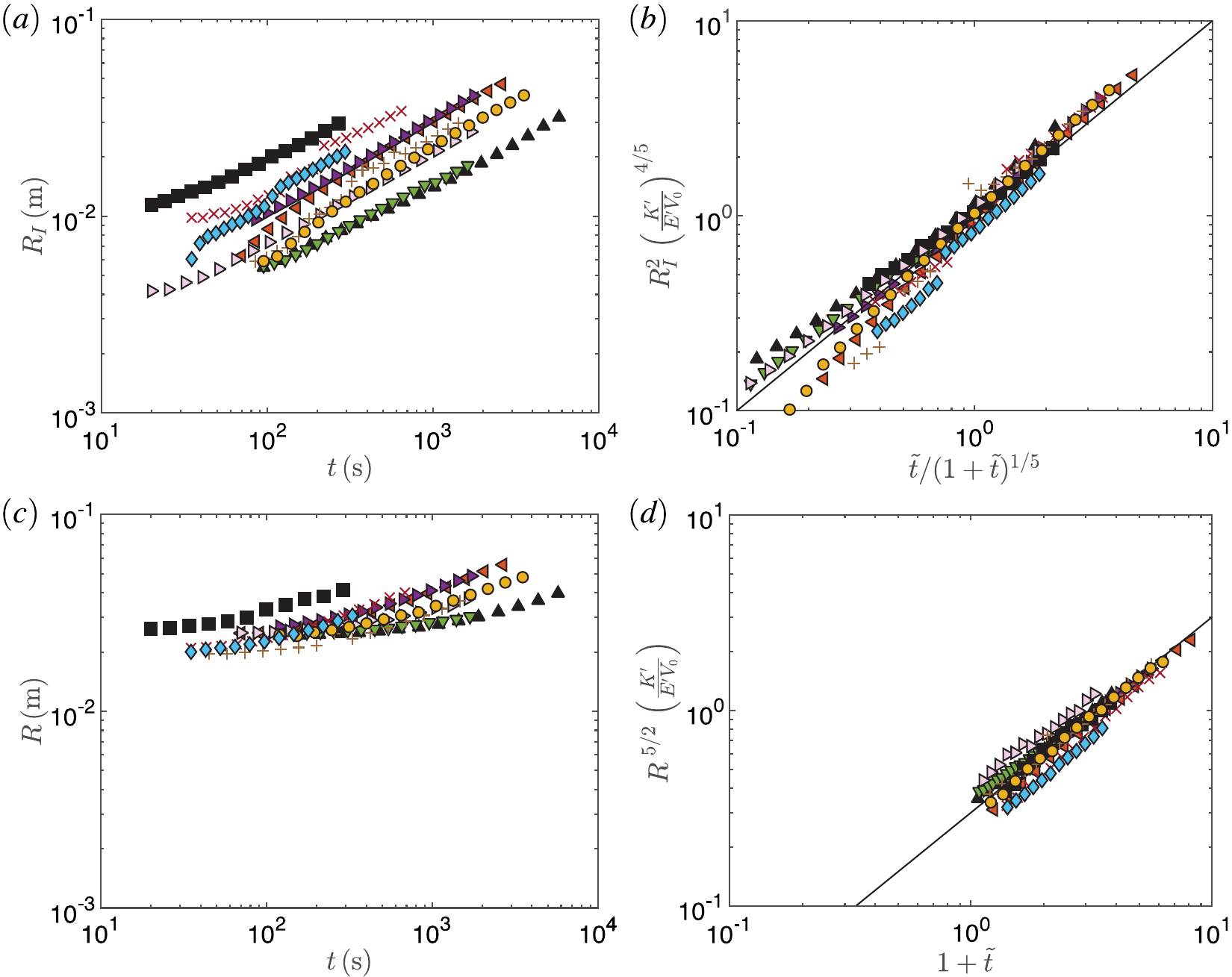}} 
\caption{Dynamics of the fracture for low-viscosity aqueous phase. (a) Dependence of the position of the interface on time for experiments 1 to 10 (see table \ref{tab:my-exp} for corresponding injection parameters). (b) Rescaled interface position as a function of rescaled time, based on the toughness scaling laws in table \ref{tab:my-table}. (c) Dependence of the fracture radius on time. (d) Rescaled radius as a function of rescaled time, based on the toughness scaling laws in table \ref{tab:my-table}. The black curves represent the best linear fit with a slope of 1.}
\label{fig:TFT}
\end{figure}

\subsection{Displacement flow}
Once the pre-fracture is formed, the valve is immediately switched to the second immiscible to avoid further propagation of the pre-fracture \citep{Mori2021}. The material properties of the gelatin contribute to the definition of the propagation regime. Low stiffness gelatin allows the observation of the toughness regime, while the toughness regime is most commonly reached in stiffer hydrogel. Thus, we characterize the displacement flow and fracture propagation in the toughness regime with the experimental systems 1 through 10 (see table \ref{tab:my-exp}), in which the pre-fracture is also formed in the toughness regime. The injected fluid is water. Similarly, we investigate the viscous regime in experiments 11 through 16. The displacing fluid is a syrup whose high viscosity is of the same magnitude as the silicone oil in the pre-fractures. 
During the displacement flow, the fracture continues its radial expansion but with a different dynamic from the one observed during the formation of the pre-fracture. To study the radial expansion of the fracture during the injection, we track the position of the interface between the two fluids $R_I$ and the radius of the fracture $R$ over time. In figure \ref{fig:TFT}, we plot the results of the experiments in the toughness regime. The radial position of the interface increases with time, similarly to what was observed for a single fluid injection. The annular region of displaced fluid between $R_I$ and $R$ moves outward and its thickness $R-R_I$ decreases over time. Using the equations in table 2, we plot the rescaled radii with respect to the relevant dimensionless time on the log-log scale. The rescaled radial positions of the interface collapses on a line of slope 1 and prefactor 1. The rescaled radii of the fracture collapses on a line of slope 1 with a prefactor of 0.3. We note that for large values of time $\tilde{t}>>1$, the fracture dynamics for the displacement flow is expected to become similar to the fracture dynamics for a single fluid. Indeed, if we plot the dimensionless radius as $R^{5/2} \frac{K'}{E'V_0}$, the prefactor of 0.3 for the displacement flow is comparable with the prefactor for a single fluid $0.7^{5/2} \approx 0.4$.
\begin{figure}
\centering
{\includegraphics[width=1\linewidth]{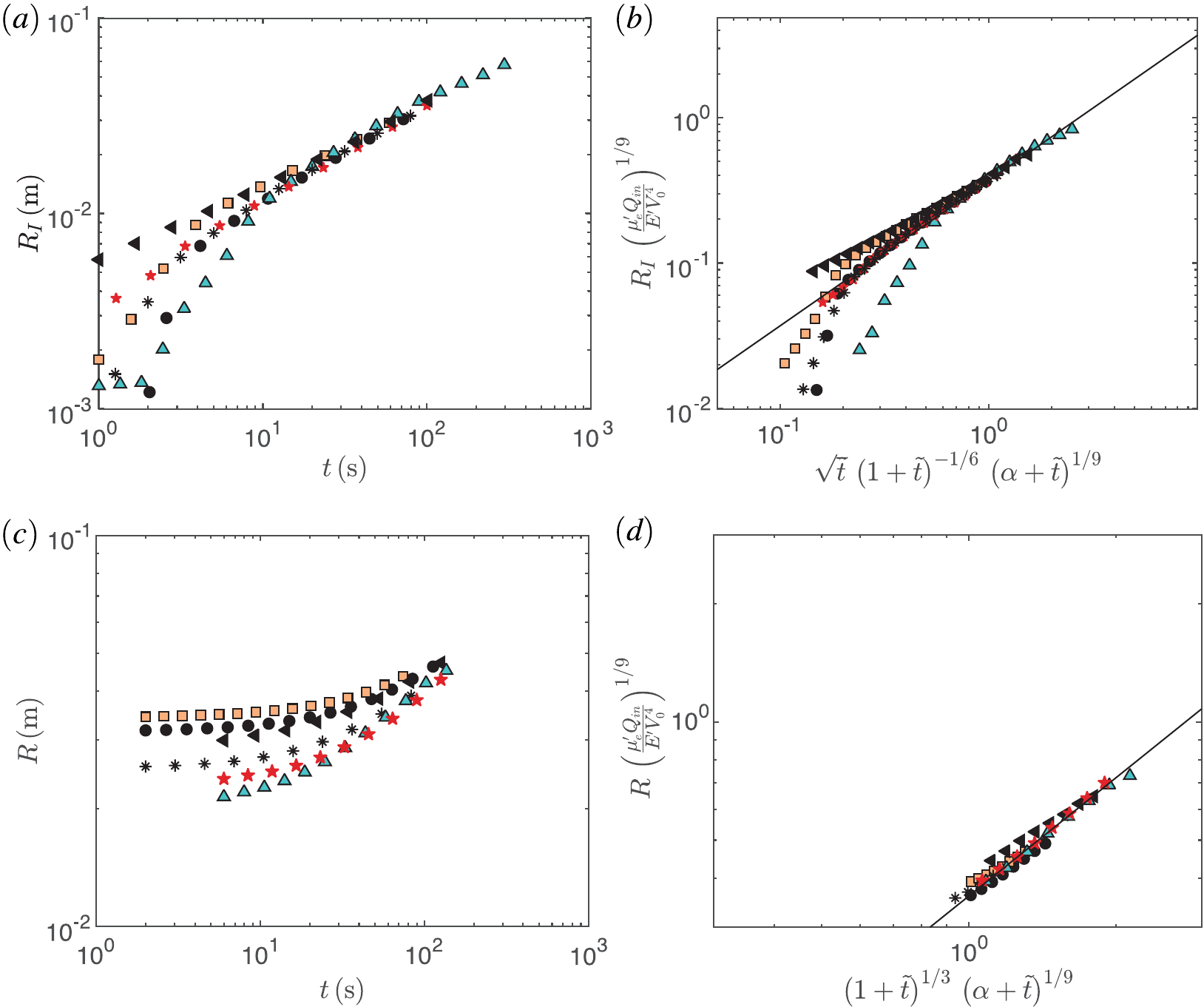}} 
\caption{Dynamics of the fracture for high-viscosity aqueous phase. (a) Dependence of the position of the interface on time for experiments 11 to 16 (see table \ref{tab:my-exp} for corresponding injection parameters). (b) Rescaled interface position as a function of rescaled time, based on the viscous scaling laws in table \ref{tab:my-table}. (c) Dependence of the fracture radius on time. (d) Rescaled radius as a function of the rescaled time, based on the viscous scaling laws in table \ref{tab:my-table}. The black curves represent the best linear fit with a slope of 1.}
\label{fig:TFV}
\end{figure}
In figure \ref{fig:TFV}, we plot the results of the displacement flow experiments in the viscous regime. The results can be rescaled using equations in the corresponding column of table 2. We plot the rescaled radii with respect to the relevant functions of the dimensionless time on the log-log scale. The rescaled radial positions of the interface collapse on a line of slope 1 and prefactor 0.37. The rescaled radii of the fracture collapses on a line of slope 1 with a prefactor of 0.36. {We note that for large values of time $\tilde{t}>>1$, the fracture dynamics for the displacement flow is expected to become similar to the fracture dynamics for a single fluid. Since at large values of $\tilde{t}$, $\left(1+\tilde{t}\right)^{1/3}\left(\alpha+\tilde{t}\right)^{1/9} \approx \, \tilde{t}^{4/9}$, the identical prefactor for the scaling of the radius for the pre-fracture and the fracture indicates that indeed the single fluid behavior is recover for large injection times. We also note that for large values of the fracture radius, the experimental data fall below the trend line. This is due to the slow down of the growth due to the confinement of the gelatin bloc. Experimentally, this is associated with the tilting of the fracture and the formation of finger-like structures.}

\begin{figure}
\centering
{\includegraphics[width=1\linewidth]{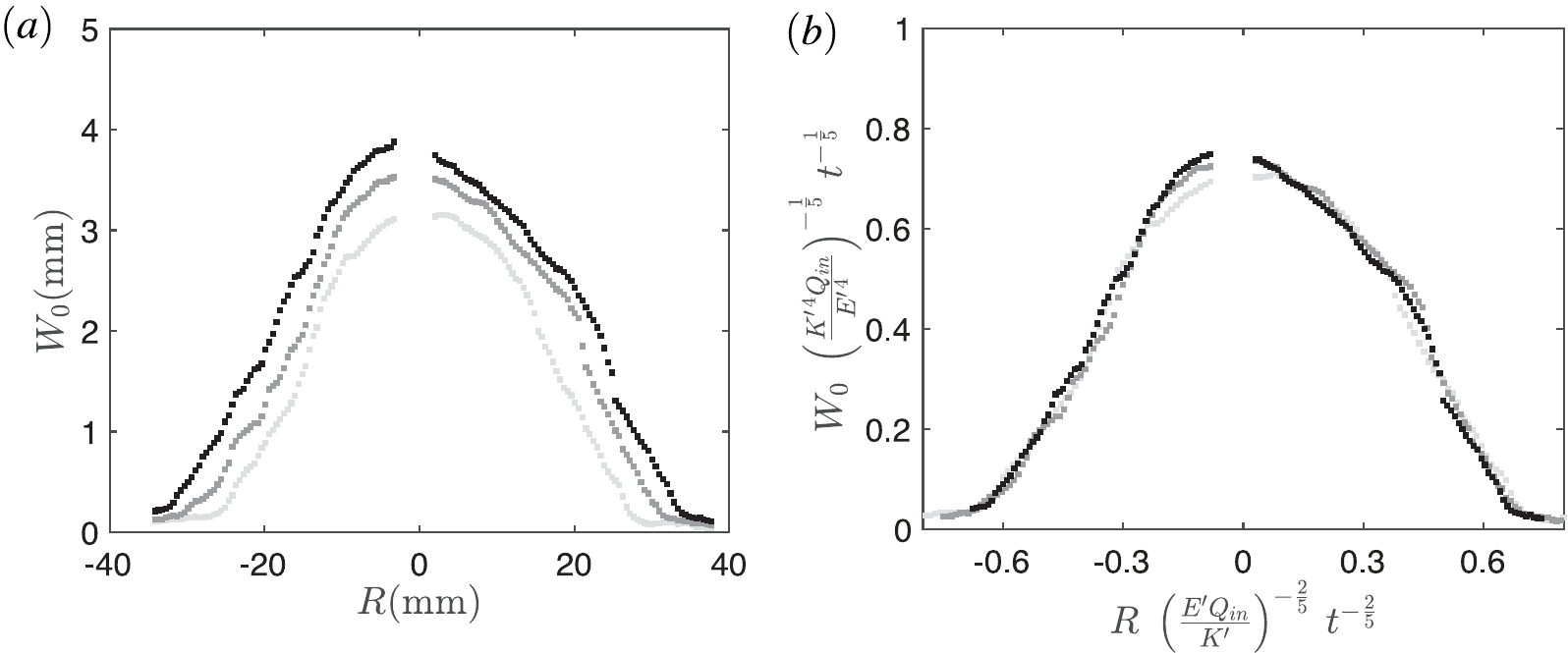}} 
\caption{Fracture profile for low-viscosity aqueous phase. (a) Fracture profiles at $t = [150,\,250,\,350]$ s with time increasing from clear to dark gray. (b) Rescaled fracture profiles using the toughness scaling laws in table \ref{tab:my-table}. Experimental parameters: $E=30$ kPa, $\mu_0= 10$ mPa.s, $Q_0 = 0.3$ ml.min$^{-1}$, $V_0 = 2.5$ ml, $\mu_{in}= 1$ mPa.s, $Q_{in} = 1.8$ ml.min$^{-1}$.}
\label{fig:TFTA}
\end{figure}

\begin{figure}[h]
\centering
{\includegraphics[width=1\linewidth]{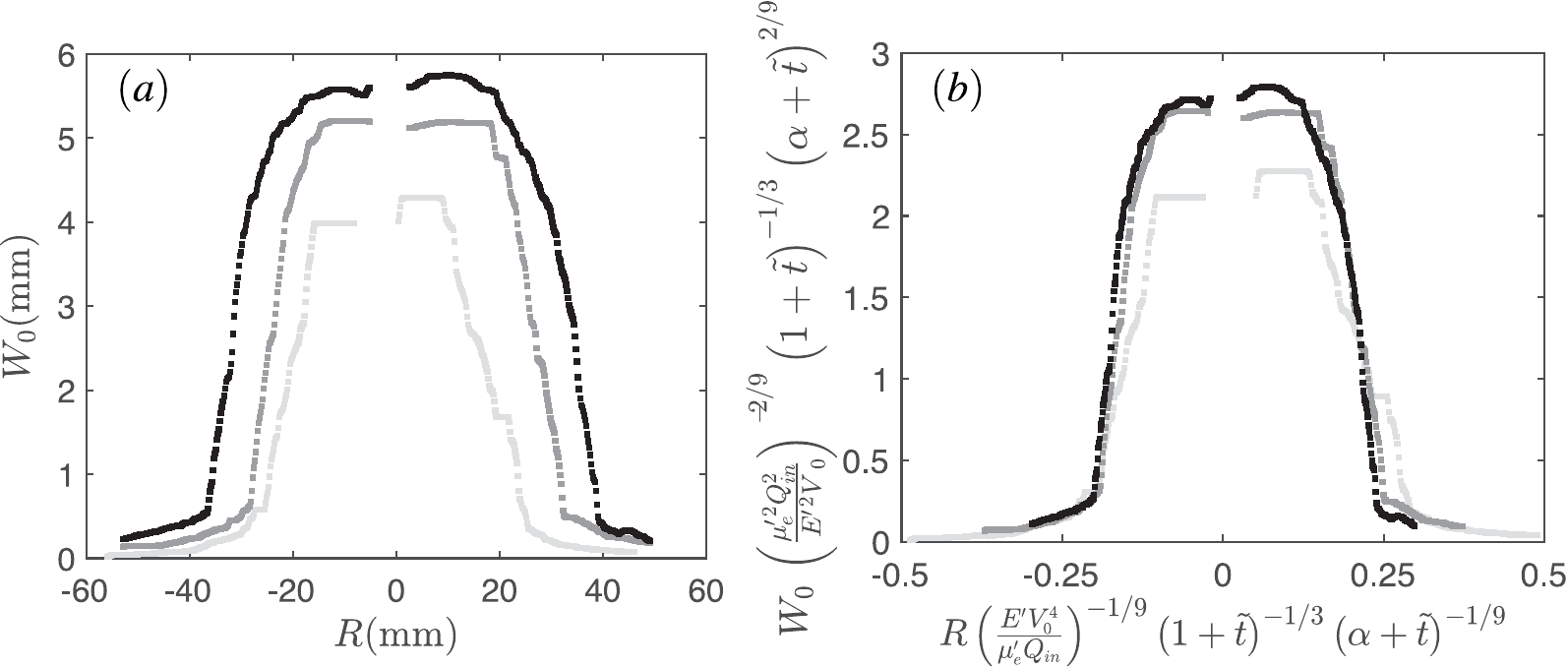}} 
\caption{Fracture profile for high-viscosity aqueous phase. (a) Fracture profiles at $t = [5,\,10,\,40]$ s with time increasing from clear to dark gray. (b) Rescaled fracture profiles using the viscous scaling laws in table \ref{tab:my-table}. Experimental parameters: $E=88$ kPa, $\mu_0= 10$ Pa.s, $Q_0 = 10$ ml.min$^{-1}$, $V_0 = 8$ ml, $\mu_{in}= 8.6$ Pa.s, $Q_{in} = 10$ ml.min$^{-1}$.}
\label{fig:TFVA}
\end{figure}

For the displacement flows, we measure the fracture aperture in the two fluids simultaneously, relying on two different dyes (see figures \ref{fig:TFTA} and \ref{fig:TFVA}). Since the two dyes absorb the light differently, the absorbance profile presents a discontinuity across the interface. The absorbance values are converted to thickness measurement using Beer's law and the calibration parameters. The experimental results plotted in figures \ref{fig:TFTA}(a) and \ref{fig:TFVA}(a) show that the aperture is a continuous function of the radial position. The aperture-radius curves are rescaled using the scalings for both the aperture and radius. In the toughness regime, both the rescaled aperture and radius collapse resulting in overlapping profiles, see figure \ref{fig:TFTA}(b). In the viscous regime, the rescaled profiles collapse on a single curve, see figure \ref{fig:TFVA}(b). The two data sets presented here are representative of all the experiments. In summary, the experimental observations establish the existence of two regimes and validate the respective scaling relations.
\section{Conclusion}

In this study, we model the properties of fractures driven by displacement flows by revisiting the theoretical framework established for single fluid injections. We derive scaling relationships for the position of the interface between the two fluids, the radius and aperture of the fracture in a brittle elastic matrix in the viscous-dominated and toughness-dominated regimes. We define a dimensionless time, which is equal to the ratio of the volume of displacing fluid injected over the volume of the pre-fracture. In the toughness regime, the propagation dynamics is independent of the fluid properties. As a result, the fracture dynamics for the displacement flow are the same as the dynamics for a single fluid, with the addition initial finite volume $V_0$.  In the viscous regime, however, the two fluids that fill the fracture contribute to the viscous dissipation. Over time, the relative volume of the two fluids changes. To describe the viscous dissipation in the fracture, we therefore define an average viscosity which accounts for the relative volume of displaced and displacing fluid. The scalings are compared to experimental results obtained by successively injecting an oil phase and an aqueous phase in a gelatin block. The experiments confirm the existence of two regimes of fracture propagation and are in good agreement with the derived relationships. \\
This study focuses on displacement flows of immiscible fluids with comparable viscosities. Other types of displacement flows are common in fracture. For example, industrial applications involve the sequential injection of miscible aqueous fluids, some of which can be complex fluids such as suspensions of particles or polymer solutions. Recent and future efforts to characterize and model multiphase flows in fractures should ultimately support efficient hydraulic fracturing operations.


\begin{thebibliography}{73}
\expandafter\ifx\csname natexlab\endcsname\relax\def\natexlab#1{#1}\fi
\def\au#1{#1} \def\ed#1{#1} \def\yr#1{#1}\def\at#1{#1}\def\jt#1{\textit{#1}} \def\bt#1{#1}\def\bvol#1{\textbf{#1}} \def\vol#1{#1} \def\pg#1{#1} \def\publ#1{#1}\def\arxiv#1{#1}\def\org#1{#1}\def\st#1{\textit{#1}}

\bibitem[Al-Housseiny \& Stone(2013)]{Al2013}
{\sc \au{Al-Housseiny, T.~T.} \& \au{Stone, H.~A.}} \yr{2013}  \at{{Controlling viscous fingering in tapered Hele-Shaw cells}}.  \jt{Phys. Fluids}  \bvol{25}~(9),  \pg{092102 -- 12}.

\bibitem[Al-Housseiny {\em et~al.\/}(2012)Al-Housseiny, Tsai \& Stone]{Al2012}
{\sc \au{Al-Housseiny, T.~T.}, \au{Tsai, P.~A.} \& \au{Stone, H.~A.}} \yr{2012}  \at{{Control of interfacial instabilities using flow geometry}}.  \jt{Nat. Phys.}  \bvol{8}~(10),  \pg{747--750}.

\bibitem[Alessi {\em et~al.\/}(2017)Alessi, Zolfaghari, Kletke, Gehman, Allen \& Goss]{Alessi2017}
{\sc \au{Alessi, D.~S.}, \au{Zolfaghari, A.}, \au{Kletke, S.}, \au{Gehman, J.}, \au{Allen, D.~M.} \& \au{Goss, G.~G.}} \yr{2017}  \at{{Comparative analysis of hydraulic fracturing wastewater practices in unconventional shale development: Water sourcing, treatment and disposal practices}}.  \jt{Can. Water Resour. J.}  \bvol{42}~(2),  \pg{1--17}.

\bibitem[Bao \& Eaton(2016)]{Bao2016}
{\sc \au{Bao, X.} \& \au{Eaton, D.~W.}} \yr{2016}  \at{{Fault activation by hydraulic fracturing in western Canada}}.  \jt{Science}  \bvol{354}~(6318),  \pg{1406--1409}.

\bibitem[Barbati {\em et~al.\/}(2016)Barbati, Desroches, Robisson \& McKinley]{Barbati2016}
{\sc \au{Barbati, A.~C.}, \au{Desroches, J.}, \au{Robisson, A.} \& \au{McKinley, G.~H.}} \yr{2016}  \at{{Complex Fluids and Hydraulic Fracturing}}.  \jt{Annu. Rev. Chem. Biomol. Eng.}  \bvol{7}~(1),  \pg{415 -- 453}.

\bibitem[Barboza {\em et~al.\/}(2021)Barboza, Chen \& Li]{Barboza2021}
{\sc \au{Barboza, B.~R.}, \au{Chen, B.} \& \au{Li, C.}} \yr{2021}  \at{{A Review on Proppant Transport Modelling}}.  \jt{J. Petrol. Sci. Eng.}  \bvol{204},  \pg{108753}.

\bibitem[Barenblatt(1956)]{Barenblatt1956}
{\sc \au{Barenblatt, G.~I.}} \yr{1956}  \at{{On the formation of horizontal cracks in hydraulic fracture of an oil-bearing stratum}}.  \jt{Prikl. Mat. Mech.}  \bvol{20},  \pg{475--486}.

\bibitem[Baumberger \& Ronsin(2020)]{Baumberger2020}
{\sc \au{Baumberger, T.} \& \au{Ronsin, O.}} \yr{2020}  \at{{Environmental control of crack propagation in polymer hydrogels}}.  \jt{Mech. Soft Mater.}  \bvol{2}~(1),  \pg{14}.

\bibitem[Bessmertnykh {\em et~al.\/}(2021)Bessmertnykh, Dontsov \& Ballarini]{Bess2021}
{\sc \au{Bessmertnykh, A.}, \au{Dontsov, E.} \& \au{Ballarini, R.}} \yr{2021}  \at{{Semi-Infinite Hydraulic Fracture Driven by a Sequence of Power-Law Fluids}}.  \jt{J. Eng. Mech.}  \bvol{147}~(10),  \pg{04021064}.

\bibitem[Bunger(2006)]{Bunger2006}
{\sc \au{Bunger, A~P}} \yr{2006}  \at{{A photometry method for measuring the opening of fluid-filled fractures}}.  \jt{Meas. Sci. Technol.}  \bvol{17}~(12),  \pg{3237}.

\bibitem[Bunger \& Detournay(2005)]{Bunger2005}
{\sc \au{Bunger, A.~P.} \& \au{Detournay, E.}} \yr{2005}  \at{{Asymptotic solution for a penny-shaped near-surface hydraulic fracture}}.  \jt{Eng. Fract. Mech.}  \bvol{72}~(16),  \pg{2468--2486}.

\bibitem[Bunger \& Detournay(2008)]{Bunger2008}
{\sc \au{Bunger, A.~P.} \& \au{Detournay, E.}} \yr{2008}  \at{{Experimental validation of the tip asymptotics for a fluid-driven crack}}.  \jt{J. Mech. Phys. Solids}  \bvol{56}~(11),  \pg{3101--3115}.

\bibitem[Bunger {\em et~al.\/}(2013)Bunger, Gordeliy \& Detournay]{Bunger2013}
{\sc \au{Bunger, A.~P.}, \au{Gordeliy, E.} \& \au{Detournay, E.}} \yr{2013}  \at{{Comparison between laboratory experiments and coupled simulations of saucer-shaped hydraulic fractures in homogeneous brittle-elastic solids}}.  \jt{J. Mech. Phys. Solids}  \bvol{61}~(7),  \pg{1636--1654}.

\bibitem[Caulk {\em et~al.\/}(2016)Caulk, Ghazanfari, Perdrial \& Perdrial]{Caulk2016}
{\sc \au{Caulk, R.~A.}, \au{Ghazanfari, E.}, \au{Perdrial, J.~N.} \& \au{Perdrial, N.}} \yr{2016}  \at{{Experimental investigation of fracture aperture and permeability change within Enhanced Geothermal Systems}}.  \jt{Geothermics}  \bvol{62},  \pg{12--21}.

\bibitem[Chen(1989)]{Chen1989}
{\sc \au{Chen, J.-D.}} \yr{1989}  \at{{Growth of radial viscous fingers in a Hele-Shaw cell}}.  \jt{J. Fluid Mech.}  \bvol{201},  \pg{223--242}.

\bibitem[Chen {\em et~al.\/}(2017)Chen, Fang, Wu \& Hu]{Chen2017}
{\sc \au{Chen, Y.‐F.}, \au{Fang, S.}, \au{Wu, D.‐S.} \& \au{Hu, R.}} \yr{2017}  \at{{Visualizing and quantifying the crossover from capillary fingering to viscous fingering in a rough fracture}}.  \jt{Water Resour. Res.}  \bvol{53}~(9),  \pg{7756--7772}.

\bibitem[Cottin {\em et~al.\/}(2010)Cottin, Bodiguel \& Colin]{Cottin2010}
{\sc \au{Cottin, C.}, \au{Bodiguel, H.} \& \au{Colin, A.}} \yr{2010}  \at{{Drainage in two-dimensional porous media: From capillary fingering to viscous flow}}.  \jt{Phys. Rev. E}  \bvol{82}~(4),  \pg{046315}.

\bibitem[Cueto-Felgueroso \& Juanes(2013)]{Cueto2013}
{\sc \au{Cueto-Felgueroso, L.} \& \au{Juanes, R.}} \yr{2013}  \at{{Forecasting long-term gas production from shale}}.  \jt{Proc. National Acad. Sci. USA}  \bvol{110}~(49),  \pg{19660--19661}.

\bibitem[Desroches {\em et~al.\/}(1994)Desroches, Detournay, Lenoach, Papanastasiou, Pearson, Thiercelin \& Cheng]{Desroches1994}
{\sc \au{Desroches, J.}, \au{Detournay, E.}, \au{Lenoach, B.}, \au{Papanastasiou, P.}, \au{Pearson, J. R.~A.}, \au{Thiercelin, M.} \& \au{Cheng, A.}} \yr{1994}  \at{{The crack tip region in hydraulic fracturing}}.  \jt{Proc. R. Soc. Lond. A}  \bvol{447}~(1929),  \pg{39--48}.

\bibitem[Detournay(2004)]{Detournay2004}
{\sc \au{Detournay, E.}} \yr{2004}  \at{{Propagation regimes of fluid-driven fractures in impermeable rocks}}.  \jt{Int. J. Geomech.}  \bvol{4}~(1),  \pg{35--45}.

\bibitem[Detournay(2016)]{Detournay2016}
{\sc \au{Detournay, E.}} \yr{2016}  \at{{Mechanics of Hydraulic Fractures}}.  \jt{Annu. Rev. Fluid Mech.}  \bvol{48}~(1),  \pg{311 -- 339}.

\bibitem[Detournay \& Garagash(2003)]{Detournay2003}
{\sc \au{Detournay, E.} \& \au{Garagash, D.~I.}} \yr{2003}  \at{{The near-tip region of a fluid-driven fracture propagating in a permeable elastic solid}}.  \jt{J. Fluid Mech.}  \bvol{494},  \pg{1--32}.

\bibitem[Detournay \& Peirce(2014)]{Detournay2014}
{\sc \au{Detournay, E.} \& \au{Peirce, A.}} \yr{2014}  \at{{On the moving boundary conditions for a hydraulic fracture}}.  \jt{Int. J. Eng. Sci.}  \bvol{84},  \pg{147--155}.

\bibitem[Garagash \& Detournay(2000)]{Garagash2000}
{\sc \au{Garagash, D.~I.} \& \au{Detournay, E.}} \yr{2000}  \at{{The tip region of a fluid-driven fracture in an elastic medium}}.  \jt{J. Appl. Mech.}  \bvol{67}~(1),  \pg{183--192}.

\bibitem[Garagash \& Detournay(2005)]{Garagash2005}
{\sc \au{Garagash, D.~I.} \& \au{Detournay, E.}} \yr{2005}  \at{{Plane-Strain Propagation of a Fluid-Driven Fracture: Small Toughness Solution}}.  \jt{J. Appl. Mech.}  \bvol{72}~(6),  \pg{916--928}.

\bibitem[Giuseppe {\em et~al.\/}(2009)Giuseppe, Funiciello, Corbi, Ranalli \& Mojoli]{Giuseppe2009}
{\sc \au{Giuseppe, E.~Di}, \au{Funiciello, F.}, \au{Corbi, F.}, \au{Ranalli, G.} \& \au{Mojoli, G.}} \yr{2009}  \at{{Gelatins as rock analogs: A systematic study of their rheological and physical properties}}.  \jt{Tectonophysics}  \bvol{473}~(3-4),  \pg{391--403}.

\bibitem[Glass {\em et~al.\/}(2003)Glass, Rajaram \& Detwiler]{Glass2003}
{\sc \au{Glass, R.~J.}, \au{Rajaram, H.} \& \au{Detwiler, R.~L.}} \yr{2003}  \at{{Immiscible displacements in rough-walled fractures: Competition between roughening by random aperture variations and smoothing by in-plane curvature}}.  \jt{Phys. Rev. E}  \bvol{68}~(6),  \pg{061110}.

\bibitem[Homsy(1987)]{Homsy1987}
{\sc \au{Homsy, G.~M.}} \yr{1987}  \at{{Viscous Fingering in Porous Media}}.  \jt{Annu. Rev. Fluid Mech.}  \bvol{19}~(1),  \pg{271--311}.

\bibitem[Hormozi \& Frigaard(2017)]{Hormozi2017}
{\sc \au{Hormozi, S.} \& \au{Frigaard, I.~A.}} \yr{2017}  \at{{Dispersion of solids in fracturing flows of yield stress fluids}}.  \jt{J. Fluid Mech.}  \bvol{830},  \pg{93 -- 137}.

\bibitem[Huppert \& Neufeld(2013)]{Huppert2013}
{\sc \au{Huppert, H.~E.} \& \au{Neufeld, J.~A.}} \yr{2013}  \at{{The Fluid Mechanics of Carbon Dioxide Sequestration}}.  \jt{Annu. Rev. Fluid Mech.}  \bvol{46}~(1),  \pg{255--272}.

\bibitem[Jia {\em et~al.\/}(2019)Jia, Tsau \& Barati]{Jia2019}
{\sc \au{Jia, B.}, \au{Tsau, J.-S.} \& \au{Barati, R.}} \yr{2019}  \at{{A review of the current progress of CO2 injection EOR and carbon storage in shale oil reservoirs}}.  \jt{Fuel}  \bvol{236},  \pg{404--427}.

\bibitem[Kanninen \& Popelar(1985)]{Kanninen1985}
{\sc \au{Kanninen, M.~F.} \& \au{Popelar, C.~H.}} \yr{1985} {\em {Advanced Fracture Mechanics}\/}. {\em Oxford Engineering Science Series\/} 15.  \publ{Oxford University Press}.

\bibitem[Kavanagh {\em et~al.\/}(2013)Kavanagh, Menand \& Daniels]{Kavanagh2013}
{\sc \au{Kavanagh, J.~L.}, \au{Menand, T.} \& \au{Daniels, K.~A.}} \yr{2013}  \at{{Gelatine as a crustal analogue: Determining elastic properties for modelling magmatic intrusions}}.  \jt{Tectonophysics}  \bvol{582},  \pg{101--111}.

\bibitem[Khristianovic \& Zheltov(1955)]{Khris1955}
{\sc \au{Khristianovic, S.~A.} \& \au{Zheltov, Y.~P.}} \yr{1955}  \at{{Formation of Vertical Fractures by Means of Highly Viscous Liquid}}.  \jt{4th World Petrol. Congr. Proc.}  \bvol{2},  \pg{576--586}.

\bibitem[Lai {\em et~al.\/}(2018)Lai, Rallabandi, Perazzo, Zheng, Smiddy \& Stone]{Lai2018}
{\sc \au{Lai, C.-Y.}, \au{Rallabandi, B.}, \au{Perazzo, A.}, \au{Zheng, Z.}, \au{Smiddy, S.~E.} \& \au{Stone, H.~A.}} \yr{2018}  \at{{Foam-driven fracture}}.  \jt{Proc. National Acad. Sci. USA}  \bvol{115}~(32),  \pg{8082 -- 8086}.

\bibitem[Lai {\em et~al.\/}(2016)Lai, Zheng, Dressaire \& Stone]{Lai2016a}
{\sc \au{Lai, C.-Y.}, \au{Zheng, Z.}, \au{Dressaire, E.} \& \au{Stone, H.~A.}} \yr{2016}  \at{{Fluid-driven cracks in an elastic matrix in the toughness-dominated limit}}.  \jt{Phil. Trans. R. Soc. Lond. A}  \bvol{374}~(2078),  \pg{20150425}.

\bibitem[Lai {\em et~al.\/}(2015)Lai, Zheng, Dressaire, Wexler \& Stone]{Lai2015}
{\sc \au{Lai, C.-Y.}, \au{Zheng, Z.}, \au{Dressaire, E.}, \au{Wexler, J.~S.} \& \au{Stone, H.~A.}} \yr{2015}  \at{{Experimental study on penny-shaped fluid-driven cracks in an elastic matrix}}.  \jt{Proc. R. Soc. Lond. A}  \bvol{471}~(2182),  \pg{20150255 -- 10}.

\bibitem[Lecampion {\em et~al.\/}(2017)Lecampion, Desroches, Jeffrey \& Bunger]{Lecampion2017}
{\sc \au{Lecampion, B.}, \au{Desroches, J.}, \au{Jeffrey, R.~G.} \& \au{Bunger, A.~P.}} \yr{2017}  \at{{Experiments versus theory for the initiation and propagation of radial hydraulic.pdf}}.  \jt{J. Geophys. Res. Solid}  \bvol{122}~(2),  \pg{1239--1263}.

\bibitem[Lenormand {\em et~al.\/}(1988)Lenormand, Touboul \& Zarcone]{Lenormand1988}
{\sc \au{Lenormand, R.}, \au{Touboul, E.} \& \au{Zarcone, C.}} \yr{1988}  \at{{Numerical models and experiments on immiscible displacements in porous media}}.  \jt{J. Fluid Mech.}  \bvol{189},  \pg{165--187}.

\bibitem[Lenormand {\em et~al.\/}(1983)Lenormand, Zarcone \& Sarr]{Lenormand1983}
{\sc \au{Lenormand, R.}, \au{Zarcone, C.} \& \au{Sarr, A.}} \yr{1983}  \at{{Mechanisms of the displacement of one fluid by another in a network of capillary ducts}}.  \jt{J. Fluid Mech.}  \bvol{135},  \pg{337--353}.

\bibitem[Lister \& Kerr(1991)]{Lister1991}
{\sc \au{Lister, J.~R.} \& \au{Kerr, R.~C.}} \yr{1991}  \at{{Fluid‐mechanical models of crack propagation and their application to magma transport in dykes}}.  \jt{J. Geophys. Res.}  \bvol{96}~(B6),  \pg{10049--10077}.

\bibitem[Lu {\em et~al.\/}(2019)Lu, Browne, Amchin, Nunes \& Datta]{Lu2019}
{\sc \au{Lu, N.~B.}, \au{Browne, C.~A.}, \au{Amchin, D.~B.}, \au{Nunes, J.~K.} \& \au{Datta, S.~S.}} \yr{2019}  \at{{Controlling capillary fingering using pore size gradients in disordered media}}.  \jt{Phys. Rev. Fluids}  \bvol{4}~(8),  \pg{084303}.

\bibitem[Luo {\em et~al.\/}(2017)Luo, Zhu, Guo, Tan, Zhuang, Liu, Zhang, Xiang \& Rohn]{Luo2017}
{\sc \au{Luo, J.}, \au{Zhu, Y.}, \au{Guo, Q.}, \au{Tan, L.}, \au{Zhuang, Y.}, \au{Liu, M.}, \au{Zhang, C.}, \au{Xiang, Wei} \& \au{Rohn, Joachim}} \yr{2017}  \at{{Experimental investigation of the hydraulic and heat-transfer properties of artificially fractured granite}}.  \jt{Sci. Rep.}  \bvol{7}~(1),  \pg{39882}.

\bibitem[Menand \& Tait(2002)]{Menand2002}
{\sc \au{Menand, T.} \& \au{Tait, S.~R.}} \yr{2002}  \at{{The propagation of a buoyant liquid‐filled fissure from a source under constant pressure: An experimental approach}}.  \jt{J. Geophys. Res.}  \bvol{107}~(B11),  \pg{ECV 16--1--ECV 16--14}.

\bibitem[Moukhtari \& Lecampion(2018)]{Moukhtari2018}
{\sc \au{Moukhtari, F.-E.} \& \au{Lecampion, B.}} \yr{2018}  \at{{A semi-infinite hydraulic fracture driven by a shear-thinning fluid}}.  \jt{J. Fluid Mech.}  \bvol{838},  \pg{573--605}.

\bibitem[Murphy {\em et~al.\/}(1981)Murphy, Tester, Grigsby \& Potter]{Murphy1981}
{\sc \au{Murphy, H.~D.}, \au{Tester, J.~W.}, \au{Grigsby, C.~O.} \& \au{Potter, R.~M.}} \yr{1981}  \at{{Energy extraction from fractured geothermal reservoirs in low‐permeability crystalline rock}}.  \jt{J. Geophys. Res.}  \bvol{86}~(B8),  \pg{7145--7158}.

\bibitem[Möri \& Lecampion(2021)]{Mori2021}
{\sc \au{Möri, A.} \& \au{Lecampion, Brice}} \yr{2021}  \at{{Arrest of a radial hydraulic fracture upon shut-in of the injection}}.  \jt{Int. J. Solids Struct.}  \bvol{219},  \pg{151--165}.

\bibitem[Osiptsov(2017)]{Osiptsov2017}
{\sc \au{Osiptsov, A.~A.}} \yr{2017}  \at{{Fluid Mechanics of Hydraulic Fracturing: a Review}}.  \jt{J. Petrol. Sci. Eng.}  \bvol{156},  \pg{513--535}.

\bibitem[O’Keeffe \& Linden(2017)]{OKeeffe2017}
{\sc \au{O’Keeffe, N.J.} \& \au{Linden, P.F.}} \yr{2017}  \at{{Hydrogel as a Medium for Fluid-Driven Fracture Study}}.  \jt{Exp. Mech.}  \bvol{57}~(9),  \pg{1 -- 11}.

\bibitem[O’Keeffe {\em et~al.\/}(2018{\natexlab{{\em a\/}}})O’Keeffe, Huppert \& Linden]{OKeeffe2018a}
{\sc \au{O’Keeffe, N.~J.}, \au{Huppert, H.~E.} \& \au{Linden, P.~F.}} \yr{2018{\natexlab{{\em a\/}}}}  \at{{Experimental exploration of fluid-driven cracks in brittle hydrogels}}.  \jt{J. Fluid Mech.}  \bvol{844},  \pg{435 -- 458}.

\bibitem[O’Keeffe {\em et~al.\/}(2018{\natexlab{{\em b\/}}})O’Keeffe, Zheng, Huppert \& Linden]{OKeeffe2018b}
{\sc \au{O’Keeffe, N.~J.}, \au{Zheng, Z.}, \au{Huppert, H.~E.} \& \au{Linden, P.~F.}} \yr{2018{\natexlab{{\em b\/}}}}  \at{{Symmetric coalescence of two hydraulic fractures}}.  \jt{Proc. National Acad. Sci. USA}  \bvol{115}~(41),  \pg{10228 -- 10232}.

\bibitem[Parisio \& Yoshioka(2020)]{Parisio2020}
{\sc \au{Parisio, F.} \& \au{Yoshioka, K.}} \yr{2020}  \at{{Modeling Fluid Reinjection Into an Enhanced Geothermal System}}.  \jt{Geophys. Res. Lett.}  \bvol{47}~(19).

\bibitem[Park \& Homsy(1984)]{Park1984}
{\sc \au{Park, C.-W.} \& \au{Homsy, G.~M.}} \yr{1984}  \at{{Two-phase displacement in Hele Shaw cells: theory}}.  \jt{J. Fluid Mech.}  \bvol{139},  \pg{291--308}.

\bibitem[Paterson(1981)]{Paterson1981}
{\sc \au{Paterson, L.}} \yr{1981}  \at{{Radial fingering in a Hele Shaw cell}}.  \jt{J. Fluid Mech.}  \bvol{113}~(-1),  \pg{513--529}.

\bibitem[Peng {\em et~al.\/}(2015)Peng, Pihler-Puzovic, Juel, Heil \& Lister]{Peng2015}
{\sc \au{Peng, G.~G.}, \au{Pihler-Puzovic, D.}, \au{Juel, A.}, \au{Heil, M.} \& \au{Lister, J.~R.}} \yr{2015}  \at{{Displacement flows under elastic membranes. Part 2. Analysis of interfacial effects}}.  \jt{J. Fluid Mech.}  \bvol{784},  \pg{512 -- 547}.

\bibitem[Pihler-Puzovic {\em et~al.\/}(2015)Pihler-Puzovic, Juel, Peng, Lister \& Heil]{Pihler2015}
{\sc \au{Pihler-Puzovic, D.}, \au{Juel, A.}, \au{Peng, G.~G.}, \au{Lister, J.~R.} \& \au{Heil, M.}} \yr{2015}  \at{{Displacement flows under elastic membranes. Part 1. Experiments and direct numerical simulations}}.  \jt{J. Fluid Mech.}  \bvol{784},  \pg{487 -- 511}.

\bibitem[Pihler-Puzović {\em et~al.\/}(2012)Pihler-Puzović, Illien, Heil \& Juel]{Pihler2012}
{\sc \au{Pihler-Puzović, D.}, \au{Illien, P.}, \au{Heil, M.} \& \au{Juel, A.}} \yr{2012}  \at{{Suppression of Complex Fingerlike Patterns at the Interface between Air and a Viscous Fluid by Elastic Membranes}}.  \jt{Phys. Rev. Lett.}  \bvol{108}~(7),  \pg{074502}.

\bibitem[Pihler-Puzović {\em et~al.\/}(2013)Pihler-Puzović, Périllat, Russell, Juel \& Heil]{Pihler2013}
{\sc \au{Pihler-Puzović, D.}, \au{Périllat, R.}, \au{Russell, M.}, \au{Juel, A.} \& \au{Heil, M.}} \yr{2013}  \at{{Modelling the suppression of viscous fingering in elastic-walled Hele-Shaw cells}}.  \jt{J. Fluid Mech.}  \bvol{731},  \pg{162--183}.

\bibitem[Primkulov {\em et~al.\/}(2019)Primkulov, Pahlavan, Fu, Zhao, MacMinn \& Juanes]{Primkulov2019}
{\sc \au{Primkulov, B.~K.}, \au{Pahlavan, A.~A.}, \au{Fu, X.}, \au{Zhao, B.}, \au{MacMinn, C.~W.} \& \au{Juanes, R.}} \yr{2019}  \at{{Signatures of fluid–fluid displacement in porous media: wettability, patterns and pressures}}.  \jt{J. Fluid Mech.}  \bvol{875},  \pg{133 -- 13}.

\bibitem[Primkulov {\em et~al.\/}(2021)Primkulov, Pahlavan, Fu, Zhao, MacMinn \& Juanes]{Primkulov2021}
{\sc \au{Primkulov, B.~K.}, \au{Pahlavan, A.~A.}, \au{Fu, X.}, \au{Zhao, B.}, \au{MacMinn, C.~W.} \& \au{Juanes, R.}} \yr{2021}  \at{{Wettability and Lenormand's diagram}}.  \jt{J. Fluid Mech.}  \bvol{923},  \pg{A34}.

\bibitem[Rice(1968)]{Rice1968}
{\sc \au{Rice, J.~R.}} \yr{1968} {\em {Mathematical analysis in the mechanics of fracture}\/}.  \publ{Academic Press, N. Y.}

\bibitem[Rubin(1995)]{Rubin1995}
{\sc \au{Rubin, A.~M.}} \yr{1995}  \at{{Propagation of Magma-Filled Cracks}}.  \jt{Annu. Rev. Earth Planet. Sci.}  \bvol{23}~(1),  \pg{287 -- 336}.

\bibitem[Saffman \& Taylor(1958)]{Saffman1958}
{\sc \au{Saffman, P.~G.} \& \au{Taylor, G.~I.}} \yr{1958}  \at{{The penetration of a fluid into a porous medium or Hele-Shaw cell containing a more viscous liquid}}.  \jt{Proc. R. Soc. Lond. A}  \bvol{245}~(1242),  \pg{312--329}.

\bibitem[Savitski \& Detournay(2002)]{Savitski2002}
{\sc \au{Savitski, A.~A.} \& \au{Detournay, E.}} \yr{2002}  \at{{Propagation of a penny-shaped fluid-driven fracture in an impermeable rock: asymptotic solutions}}.  \jt{Int. J. Solids Struct.}  \bvol{39}~(26),  \pg{6311 -- 6337}.

\bibitem[Sneddon \& Mott(1946)]{sneddon1946}
{\sc \au{Sneddon, I.~N.} \& \au{Mott, N.~F.}} \yr{1946}  \at{{The distribution of stress in the neighbourhood of a crack in an elastic solid}}.  \jt{Proc. R. Soc. Lond. A}  \bvol{187}~(1009),  \pg{229 -- 260}.

\bibitem[Spence \& Sharp(1985)]{Spence1985}
{\sc \au{Spence, D.~A.} \& \au{Sharp, P.}} \yr{1985}  \at{{Self-similar solutions for elastohydrodynamic cavity flow}}.  \jt{Proc. R. Soc. Lond. A}  \bvol{400}~(1819),  \pg{289 -- 313}.

\bibitem[Stokes {\em et~al.\/}(1986)Stokes, Weitz, Gollub, Dougherty, Robbins, Chaikin \& Lindsay]{Stokes1986}
{\sc \au{Stokes, J.~P.}, \au{Weitz, D.~A.}, \au{Gollub, J.~P.}, \au{Dougherty, A.}, \au{Robbins, M.~O.}, \au{Chaikin, P.~M.} \& \au{Lindsay, H.~M.}} \yr{1986}  \at{{Interfacial Stability of Immiscible Displacement in a Porous Medium}}.  \jt{Phys. Rev. Lett.}  \bvol{57}~(14),  \pg{1718--1721}.

\bibitem[Tabeling {\em et~al.\/}(1987)Tabeling, Zocchi \& Libchaber]{Tabeling1987}
{\sc \au{Tabeling, P.}, \au{Zocchi, G.} \& \au{Libchaber, A.}} \yr{1987}  \at{{An experimental study of the Saffman-Taylor instability}}.  \jt{J. Fluid Mech.}  \bvol{177},  \pg{67--82}.

\bibitem[Takada(1990)]{Takada1990}
{\sc \au{Takada, A.}} \yr{1990}  \at{{Experimental study on propagation of liquid‐filled crack in gelatin: Shape and velocity in hydrostatic stress condition}}.  \jt{J. Geophys. Res.}  \bvol{95}~(B6),  \pg{8471--8481}.

\bibitem[Tanveer(1993)]{Tanveer1993}
{\sc \au{Tanveer, S.}} \yr{1993}  \at{{Evolution of Hele-Shaw interface for small surface tension}}.  \jt{J. Geophys. Res.}  \bvol{343}~(1668),  \pg{155--204}.

\bibitem[Wang {\em et~al.\/}(2018)Wang, Elsworth \& Denison]{Wang2018}
{\sc \au{Wang, J.}, \au{Elsworth, D.} \& \au{Denison, M.~K.}} \yr{2018}  \at{{Propagation, proppant transport and the evolution of transport properties of hydraulic fractures}}.  \jt{J. Fluid Mech.}  \bvol{855},  \pg{503--534}.

\bibitem[Yang {\em et~al.\/}(2019)Yang, Méheust, Neuweiler, Hu, Niemi \& Chen]{Yang2019}
{\sc \au{Yang, Z.}, \au{Méheust, Y.}, \au{Neuweiler, I.}, \au{Hu, R.}, \au{Niemi, A.} \& \au{Chen, Y.‐F.}} \yr{2019}  \at{{Modeling Immiscible Two‐Phase Flow in Rough Fractures From Capillary to Viscous Fingering}}.  \jt{Water Resour. Res.}  \bvol{55}~(3),  \pg{2033--2056}.

\bibitem[Zhao {\em et~al.\/}(2016)Zhao, MacMinn \& Juanes]{Zhao2016}
{\sc \au{Zhao, Benzhong}, \au{MacMinn, Christopher~W.} \& \au{Juanes, Ruben}} \yr{2016}  \at{{Wettability control on multiphase flow in patterned microfluidics}}.  \jt{Proc. National Acad. Sci. USA}  \bvol{113}~(37),  \pg{10251--10256}.

\end{thebibliography}
\end{document}